\documentclass[fleqn,usenatbib]{mnras}

\usepackage[T1]{fontenc}
\usepackage{ae,aecompl}

\usepackage{graphicx}	
\usepackage{amsmath}	
\usepackage{amssymb}	
\usepackage{booktabs,caption}
\usepackage{threeparttable}
\usepackage{xcolor}
\hypersetup{draft}
\usepackage[normalem]{ulem}
\usepackage[version=4]{mhchem}

\title[Dust-gas chemistry in AGB outflows]{Chemical modelling of dust-gas chemistry within AGB outflows \\ I. Effect on the gas-phase chemistry}

\author[Van de Sande et al.]{
M. Van de Sande$^{1}$\thanks{E-mail: marie.vandesande@kuleuven.be}\thanks{Postdoctoral Fellow of the Fund for Scientific Research (FWO), Flanders, Belgium},
C. Walsh$^{2}$,
T. P. Mangan$^{3}$,
L. Decin$^{1,3}$
\\
$^{1}$Department of Physics and Astronomy, Institute of Astronomy, KU Leuven, Celestijnenlaan 200D, 3001 Leuven, Belgium\\
$^{2}$School of Physics and Astronomy, University of Leeds, Leeds LS2 9JT, UK\\
$^{3}$School of Chemistry, University of Leeds, Leeds LS2 9JT, UK\\
}

\date{Accepted 2019 September 20. Received 2019 September 20; in original form 2019 August 16 }

\pubyear{2019}

\begin{document}
\label{firstpage}
\pagerange{\pageref{firstpage}--\pageref{lastpage}}
\maketitle

\begin{abstract}
Chemical modelling of AGB outflows is typically focused on either non-thermodynamic equilibrium chemistry in the inner region or photon-driven chemistry in the outer region. 
We include, for the first time, a comprehensive dust-gas chemistry in our AGB outflow chemical kinetics model, including both dust-gas interactions and grain-surface chemistry. 
The dust is assumed to have formed in the inner region, and follows an interstellar-like dust-size distribution. 
Using radiative transfer modelling, we obtain dust temperature profiles for different dust types in an O-rich and a C-rich outflow. 
We calculate a grid of models, sampling different outflow densities, drift velocities between the dust and gas, and dust types. 
Dust-gas chemistry can significantly affect the gas-phase composition, depleting parent and daughter species and increasing the abundance of certain daughter species via grain-surface formation followed by desorption/sputtering. 
Its influence depends on four factors: outflow density, dust temperature, initial composition, and drift velocity. 
The largest effects are for higher density outflows with cold dust and O-rich parent species, as these species generally have a larger binding energy. 
At drift velocities larger than $\sim 10$ km s$^{-1}$, ice mantles undergo sputtering; however, they are not fully destroyed. 
Models with dust-gas chemistry can better reproduce the observed depletion of species in O-rich outflows. 
When including colder dust in the C-rich outflows and adjusting the binding energy of CS, the depletion in C-rich outflows is also better reproduced.  
To best interpret high-resolution molecular line observations from AGB outflows, dust-gas interactions are needed in chemical kinetics models.
\end{abstract}




\section{Introduction}

During the asymptotic giant branch (AGB) phase, stars with an initial mass of up to 8 M$_\odot$ lose their outer stellar layers by means of a stellar outflow or wind. 
This creates an extended circumstellar envelope (CSE). 
CSEs are rich astrochemical laboratories, with over 100 molecules detected therein as well as newly formed dust species. 
Hence, they are important contributors to the chemical enrichment of the interstellar medium (ISM) of galaxies.
The chemical content of the CSE is determined by the elemental carbon-to-oxygen ratio (C/O) of the AGB star itself, with C/O < 1 leading to an oxygen-rich outflow and C/O > 1 leading to a carbon-rich outflow (see, e.g., \citeauthor{Habing2003}~\citeyear{Habing2003}, and references therein).

Chemistry and dynamics are closely coupled throughout the CSE. 
Close to the star, in the inner wind, chemistry is taken out of thermodynamic equilibrium (TE) because of shocks caused by the pulsating AGB star. 
In the intermediate wind, solid state dust is able to condense from the gas phase, launching a dust-driven wind. 
The material reaches its terminal velocity in the outer wind, where the chemistry is dominated by photodissociation, because interstellar UV photons can more easily penetrate this lower density region.
Hence, studying the chemistry throughout the CSE leads to a better understanding of its dynamics, because the abundances of gas-phase species are very sensitive to the physical conditions within the outflow.
Investigating the chemistry can help us understand the elusive launching mechanism in O-rich outflows, since we still do not know the mechanism behind O-rich dust growth.
However, through a combination of chemical and radiative transfer modelling, the composition of the first seed particles can be deduced \citep{Woitke2006,Hofner2008,Decin2017}.

Chemical models of CSEs are typically divided into two main types, dealing with either non-TE chemistry in the inner wind and studying the first steps of dust formation \citep[e.g.,][]{Koehler1997,Cherchneff2006,Goumans2012,Gobrecht2016,Boulangier2019}, or with photon-dominated chemistry in the outer wind \citep[e.g.,][]{Huggins1982,Nejad1984,Millar1994,Li2016}.
Despite the wealth of models developed to date, there remain persistent disagreements between observations and model predictions.
Observations show that the abundances of species such as SiO and SiS decreases in the intermediate wind of high-mass loss rate O-rich outflows, before the onset of photodissociation \citep[e.g.,][]{Bujarrabal1989,Decin2010,Verbena2019}, while H$_2$O ice has been detected around OH/IR stars \citep{Sylvester1999}.
For both O-rich and C-rich stars, general trends with mass-loss rate have been observed for certain species, where their abundance decreases with increasing mass-loss rate, hinting towards depletion of gas-phase material onto dust grains \citep[e.g., SiO, SiS, CS;][]{GonzalezDelgado2003,Massalkhi2019}. 
Despite the presence of dust grains in the CSE, most of the described models include gas-phase processes only.
Hence, the chemistry of the intermediate wind, where both gas and dust are present at high densities, and where depletion via accretion onto dust grains could occur, has not been studied so far.
Some efforts have been made, but these were restricted to a dynamical description of the formation of water ice, motivated by its detection and possible effects on the gas-phase water in the molecular envelope \citep{Jura1985,Charnley1993,Dijkstra2003,Dijkstra2006}.

Here, for the first time, we model the gas and ice chemistry of an AGB outflow using a chemical model that includes a comprehensive dust-gas chemistry.
Our model includes dust-gas interactions (accretion and thermal and non-thermal desorption) and surface reactions on dust grains, in addition to the standard gas-phase chemistry.
We investigate the effect of dust-gas chemistry on the gas-phase composition in both an O-rich and C-rich CSE.
The aim is to quantify the influence of dust-gas interactions on the abundance and distribution of observed species in the CSE and to determine if such processes can reconcile the current disagreement between gas-phase-only models and observations.
In forthcoming papers, we will explore the influence of the specific grain size distribution and the composition of the ice mantle.

The chemical kinetics model is described in Sect. \ref{sect:model}, explaining in detail the different reactions added to the chemical network and the inclusion of the grain size distribution.
Our results on the influence on the gas-phase chemistry are shown in Sect.~\ref{sect:results} for both an O-rich and a C-rich outflow, followed by the discussion and conclusions in Sects.~\ref{sect:discussion} and \ref{sect:conclusion}, respectively.

\begin{table}
	\caption{Physical parameters of the grid of chemical models.}
	\resizebox{1.0\columnwidth}{!}{%
	\label{table:modelparams}
	\begin{tabular}{ll} 
		\hline
    Mass-loss rate, $\dot{M}$        &    $10^{-5}, 10^{-6}, 10^{-7}\ \mathrm{M}_\odot\ \mathrm{yr}^{-1}$ \\
    Stellar temperature, $T_*$        & 2000 K \\
    Outflow velocity, $v_\infty$   & 5, 15 km s$^{-1}$  \\
    Stellar radius, $R_*$             & 5 $\times 10^{13}$ cm \\
    Initial radius of the model		& $10^{15}$ cm \\
    Final radius of the model		& $10^{18}$ cm \\
    Exponent temperature power-law, $\epsilon$                       & 0.7 \\
    Drift velocity, $v_\mathrm{drift}$	& 0, 5, 10, 15, 20 km s$^{-1}$  \\
		\hline
	\end{tabular}
	}
\end{table}

\begin{table}
	\caption{Parent species and their initial abundances relative to H$_2$ for the C-rich and O-rich CSE. Adopted from \citet{Agundez2010}.
} 
    \begin{tabular}{l r c c l r c}
    \hline  
    \multicolumn{3}{c}{Carbon-rich} && \multicolumn{3}{c}{Oxygen-rich}  \\  
    \cline{1-3} \cline{5-7} 
    \noalign{\smallskip}
    Species & Abun. & Ref. & & Species & Abun. & Ref. \\
    \hline
    He            & 0.17                &        &    & He        & 0.17                &     \\
    CO            & $8.0\times10^{-4}$    & (1)    &    & CO        & $3.0\times10^{-4}$    & (1) \\
    N$_2$         & $4.0\times10^{-5}$    & (2)    &    & N$_2$     & $4.0\times10^{-5}$    & (2) \\
    C$_2$H$_2$    & $8.0\times10^{-5}$    & (3)    &    & H$_2$O    & $3.0\times10^{-4}$    & (6)  \\
    HCN           & $2.0\times10^{-5}$    & (3)    &    & CO$_2$    & $3.0\times10^{-7}$    & (7)  \\
    SiO           & $1.2\times10^{-7}$  & (3)    &    & SiO       & $5.0\times10^{-5}$  & (8)  \\
    SiS           & $1.0\times10^{-6}$    & (3)    &    & SiS       & $2.7\times10^{-7}$  & (9)  \\
    CS            & $5.0\times10^{-7}$    & (3)    &    & SO        & $1.0\times10^{-6}$    & (10)  \\
    SiC$_2$       & $5.0\times10^{-8}$    & (3)    &    & H$_2$S    & $7.0\times10^{-8}$    & (11)  \\
    HCP           & $2.5\times10^{-8}$  & (3)    &    & PO        & $9.0\times10^{-8}$    & (12)  \\
    NH$_3$        & $2.0\times10^{-6}$  & (4)    &    & HCN       & $2.0\times10^{-7}$  & (13)  \\
    H$_2$O        & $1.0\times10^{-7}$  &  (5)   &    & NH$_3$    & $1.0\times10^{-7}$  & (14) \\
    \hline 
    \end{tabular}%
    \footnotesize
    { {{References.}} (1) \citet{Teyssier2006}; (2) TE abundance \citep{Agundez2010}; (3) \citet{Agundez2009};
    (4) \citet{Agundez2012}; (5) \citet{Decin2010b}; (6) \citet{Maercker2008}; (7) \citet{Tsuji1997};
    (8) \citet{Schoier2004}; (9) \citet{Schoier2007}; (10) \citet{Bujarrabal1994};
    (11) \citet{Ziurys2007}; (12) \citet{Tenenbaum2007}; (13) \citet{Decin2010}; (14) \citet{Wong2018}.
    }
    
    \label{table:Model-Parents}    
\end{table}

\section{The chemical model }     \label{sect:model}

The chemical kinetics model and the reaction network used are based on the publicly available UMIST Database for Astrochemistry (UDfA) CSE model and \textsc{Rate12} reaction network \citep{McElroy2013}\footnote{\url{http://udfa.ajmarkwick.net/index.php?mode=downloads}}.
The one-dimensional model describes a uniformly expanding outflow with constant mass-loss rate and outflow velocity, leading to a number density profile, $n(r)$, that falls as $1/r^2$, with $r$ the distance from the centre of the star.
The effects of CO self-shielding are taken into account by using a single-band approach \citep{Morris1983}, while H$_2$ is assumed to be fully self-shielded, so that $n_{\mathrm{H}_2} \approx n(r)$. 
As in \citet{VandeSande2018}, the temperature distribution is described by a power law,
\begin{equation}
    T(r) = T_* \left(\frac{r}{R_*}\right)^{-\epsilon},
\end{equation}
where $T_*$ and $R_*$ are the stellar temperature and radius, respectively. 

In this work, we extend the gas-phase only model to include dust-gas interactions and grain-surface reactions, following the recipes for models using the mean-field rate-equation approach described by \citet{Cuppen2017}.
These methods are regularly used in other astrochemical laboratories such as protoplanetary disks \citep[e.g.,][]{Walsh2014}.
All neutral species are allowed to adsorb onto the dust surface and form an ice mantle, to react on the dust surface, and to return to the gas phase through thermal desorption, photodesorption and non-thermal sputtering.
This divides the chemistry into two components: the gas-phase species and the solid-phase ice mantles covering the dust grains.
Dust grains are assumed to be present throughout the outflow and to follow the grain size distribution of the ISM \citep{Mathis1977}, where we assume that the dust size distribution of the ISM is already set in the AGB wind. 
Hence, the models start at $10^{15}$ cm $\sim$ 20 R$_*$, i.e., the dust is assumed to have formed in the inner region within 20 R$_*$.

To study the effects of including these reactions, we calculate a grid of models for both a C-rich and an O-rich outflow.
The physical parameters for the grid of models are given in Table \ref{table:modelparams}.
The parent species' initial abundances, are listed in Table \ref{table:Model-Parents}. Parent species are those formed near the photosphere of the star and are thus assumed to be present at the start of the model.
The grid samples a range in both outflow densities and drift velocities, $v_\mathrm{drift}$, between the dusty and gaseous components, where $v_\mathrm{drift} = v_\mathrm{dust} - v_\mathrm{gas}$.

In Sect.~\ref{subsect:model:chem}, the different dust-gas interactions and grain-surface reactions added to the chemical model are described.
Sect.~\ref{subsect:model:dust} describes the dust-grain size distribution and the dust temperature profile throughout the outflow.

\subsection{Chemistry}                  \label{subsect:model:chem}

The chemical reactions included in the model are divided into two categories: dust-gas interactions (Sect. \ref{subsubsect:model:chem:dustgas}) and reactions on the grain surface (Sect. \ref{subsubsect:model:chem:grainsurface}).
The reaction rates are calculated as described in \citet{Cuppen2017} and are summarised below.

\subsubsection{Dust-gas interactions}         \label{subsubsect:model:chem:dustgas}

Five types of dust-gas interactions are added to the network: accretion, thermal desorption, photodesorption, non-thermal sputtering, and cation-grain recombination.

\paragraph*{Accretion}            \label{par:model:chem:dustgas:accretion}

The rate of accretion of gas-phase species onto the dust is calculated as
\begin{equation}        \label{eq:kaccr}
    k_\mathrm{accr} = S \ \langle v \rangle \ \sigma_\mathrm{dust} \ F(Q)\ \ \ \mathrm{s}^{-1},
\end{equation}
where $S$ is the sticking coefficient of the gas-phase species, $\langle v \rangle$ is the average gas-phase thermal velocity [cm s$^{-1}$], which depends on the temperature of the gas, and $\sigma_\mathrm{dust}$ is the effective dust surface area per unit volume [cm$^{-1}$].
The sticking coefficient is assumed to be equal to one for all species except H, for which it is assumed to be 0.3, as used in the UDfA CSE model \citep{McElroy2013}.

The function $F(Q)$ describes the variation of the accretion rate with the dust drift velocity. Non-zero drift velocities lead to a larger accretion rate, because the dust grains sweep up gas-phase material. The variation is given by
\begin{equation}
    F(Q) = \frac{1}{\sqrt{\pi} Q} \left( \frac{\sqrt{\pi}}{2} \left( 2Q^2+1\right) \Phi(Q) + Q\mathrm{e}^{-Q^2}    \right),
\end{equation}
where $Q$ is the ratio between the drift velocity and thermal gas velocity and $\Phi(Q)$ is the Gaussian error integral \citep{Gail2013}.

\paragraph*{Thermal desorption}            \label{par:model:chem:dustgas:td}

The thermal desorption rate is calculated as
\begin{equation}        \label{eq:ktd}
    k_\mathrm{TD} = \nu\ \mathrm{exp}\left(-\frac{E_\mathrm{bind}}{T_\mathrm{dust}}\right)\ \ \ \mathrm{s}^{-1},
\end{equation}
where $\nu$ is a characteristic attempt frequency of the ice species to escape off the surface [s$^{-1}$], $E_\mathrm{bind}$ is the binding energy of the species to the surface [K], and $T_\mathrm{dust}$ is the dust grain temperature [K].
The attempt frequency is approximated by $10^{12}$ s$^{-1}$ \citep{Tielens1987}.
The binding energies of species to the grain surfaces are those of \textsc{Rate12} with updates from \citet{Penteado2017}.
The dust temperature is calculated as described in Sect. \ref{subsubsect:model:dust:dusttemp}.
Following \citet{Cuppen2017}, we have adjusted the zeroth-order desorption rate to a first-order rate for the sub-monolayer regime, fixing the number of active monolayers to two.

\paragraph*{Photodesorption}            \label{par:model:chem:dustgas:pd}

The photodesorption rate is calculated as
\begin{equation}        \label{eq:kpd}
    k_\mathrm{PD} = \zeta\ Y_{pd}\ F_{UV}\ \sigma_\mathrm{dust}\ \ \ \mathrm{s}^{-1},
\end{equation}
where $Y_{pd}$ is the photodesorption yield [photon$^{-1}$], $F_{UV}$ is the flux of UV photons [photons cm$^{-2}$ s$^{-1}$], and $\sigma	_\mathrm{dust}$ is the average dust grain cross section per unit volume [cm$^{-1}$].
The factor $\zeta$ [cm$^3$] takes into consideration the surface coverage of the dust and is calculated as the inverse of the total ice abundance. In the sub-monolayer regime, it is $2 \times N_s$, with two being the number of active monolayers and $N_s$ the number density of grain surface sites [cm$^{-3}$]. 
The photodesorption yield $Y_{pd}$ is taken to be equal to $10^{-3}$ photon$^{-1}$ for all ice species \citep{Oberg2016}.
The average dust grain cross section per unit volume $\sigma_\mathrm{dust}$ depends on the dust-grain size distribution and is calculated as described in Sect. \ref{subsubsect:model:dust:gsd}.
The flux of UV photons $F_{UV}$ takes into account the secondary photons produced by cosmic-ray excitation of H$_2$ as well as those from the interstellar radiation field. 
The effects of a sub-monolayer regime are taken into account when calculating the grain surface coverage factor.

\paragraph*{Sputtering}            \label{par:model:chem:dustgas:sputtering}

Collisions between dust grains and gas particles can mechanically remove the dust surface layer through (non-thermal) sputtering.
Sputtering can only occur if the energy of the incident particle is larger than the threshold energy, $E_\mathrm{th}$.
The threshold energy depends on the mass ratio between the target and the projectile and the binding energy of the target \citep{Woitke1993,Tielens1994}.
The threshold energy of, e.g., \ce{H2} sputtering of CO is $E_\mathrm{th} = 0.39$ eV, that for SiO is $E_\mathrm{th} = 2.18$ eV, and that for H$_2$O is $E_\mathrm{th} = 1.68$ eV.

\citet{Tielens1994} describe the total sputtering yield, $Y$, for normal incident projectiles as
\begin{equation}		\label{eq:yieldtielens}
    Y(E) = 4.2 \times 10^{14}\ \frac{\alpha\ S_n(E)}{U_0}\ \ \ \mathrm{species\ per\ collision},
\end{equation}
where $E$ is the energy of the impacting projectile, $\alpha$ is an energy independent function of the mass ratio between the ice target and gas-phase projectile, $S_n(E)$ is the nuclear stopping cross-section [erg cm$^{-2}$], and $U_0$ is the surface binding energy [eV] of the target. 
The surface binding energy is taken to be equal to the target's binding energy.
For the material-dependent parameter, $K$, which takes into account the mean penetrated path length of the projectile in the calculation of $\alpha$, we adopted a value of 0.01, because this provided a good fit to the experimental data by \citet{Fama2008} of He$^+$ sputtering of H$_2$O ice.
For the detailed derivation of Eq. \ref{eq:yieldtielens}, we refer to \citet{Tielens1994}.
\citet{Woitke1993} give a different description of the total sputtering yield. 
In Section \ref{subsect:discussion:comparison}, we compare both descriptions.

For energies larger than $E_\mathrm{th}$, the sputtering rate is given by
\begin{equation}		\label{eq:ksput}
    k_\mathrm{sput} = 2\ Y(E)\ k_\mathrm{accr}\ \ \ \mathrm{s}^{-1},
\end{equation}
where the factor of 2 takes non-normal incidence into account.
All details can be found in \citet{Tielens1994}.

\begin{figure}
 \includegraphics[width=0.9\columnwidth]{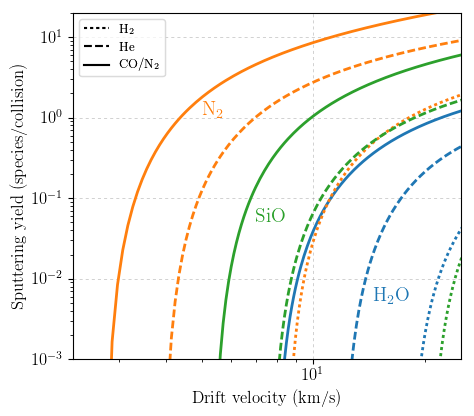}
 \caption{Sputtering yield of H$_2$O ice species (blue), N$_2$ ice species (orange), and SiO ice species (green) by collisions with different particles.
 Dotted lines: collisions with H$_2$. Dashed lines: collisions with He. Solid lines: collisions with CO or N$_2$.
  }
 \label{fig:sputteringyields}
\end{figure}

We include sputtering by H$_2$, He, CO, and N$_2$, which are the most abundant species within the outflow. 
The sputtering yield depends on the target on the grain surface, the mass of the incident projectile, and their relative drift velocity, as shown in Fig. \ref{fig:sputteringyields}.

Chemisputtering, i.e., the removal of ice species from the dust surface through chemical reactions with other species, and shattering through grain-grain collisions are ignored.
We also assume that the dust grains themselves cannot be destroyed through sputtering, because the physical bonds within the ice layer are weaker than the chemical bonds within the dust grains.
Using the binding energies for silicate and carbonaceous grains as listed by \citet{Tielens1994}, 5.7 and 4 eV respectively, we find that the sputtering by H$_2$ and He only becomes significant for drift velocities larger than $\sim$ 40 km s$^{-1}$.
In comparison, the onset of sputtering by CO and N$_2$ lies at $v_\mathrm{drift} \sim$ 20 km s$^{-1}$, which is the upper end of our considered range in drift velocities.

\paragraph*{Cation-grain recombination}            \label{par:model:chem:dustgas:cationgrain}

Because the dust grains are negatively charged, collisions between the dust and cations can lead to dissociative recombination. 
Assuming fast conduction of the electron on the grain, the cation-grain recombination rates is calculated as
\begin{equation}
\begin{split}
	k_\mathrm{CG} = k_\mathrm{accr}\ n_\mathrm{grain} & \left( 1 + \frac{e^2}{k\ a_\mathrm{grain}\ T_\mathrm{gas}} \right) \\ 
	&\left( 1 + \sqrt{\frac{2e^2}{k\ a_\mathrm{grain}\ T_\mathrm{gas} + 2e^2}} \right),
\end{split}
\end{equation}
where $e$ is the electron charge [statcoulomb]. The last two factors express the rate enhancement due to electrostatic attraction \citep{Draine1987}.

Cation-grain recombination does not play a large role in CSE chemistry, because dust grains are not the main carriers of negative charge.
Ionisation by cosmic rays and interstellar radiation give rise to a large gas-phase ionisation fraction throughout the outflow, so that electrons are the main negative charge carriers, with abundances of $\sim 10^{-8}$ around $10^{16}$ cm, rising to $\sim 10^{-4}$ relative to H$_2$ in the outer wind.
Although dissociative recombination is more important, we include cation-grain recombination for completeness of the model.

\subsubsection{Grain-surface reactions}      \label{subsubsect:model:chem:grainsurface}

On the grain surface, chemical reactions involving ice species can occur through either the diffusive Langmuir-Hinshelwood (LH) mechanism or the stick-and-hit Eley-Rideal (ER) mechanism. 
We do not include the hot-atom mechanism, because the thermalisation time scale tends to be significantly shorter than the chemical time scales.
Tunneling through barriers, both for reactions and for diffusion, is not considered here. Tunneling through barriers to diffusion is only important in the very cold regime (T $< 15$ K) where the thermal hopping rate of light atoms is negligible \citep{Cuppen2017}. This temperature is reached at the very outermost edge of the outflow only, where the chemistry is dominated by the interstellar UV radiation field. It is expected that the surface chemistry is dominated by barrierless reactions; hence, we also neglect tunneling through reaction barriers. However, this should be tested, and we will explore its impact on the ice composition in future work.

In this pilot study, we have included a simple network involving hydrogenation reactions, atom addition reactions and radical recombination reactions.
All grain-surface reactions included in our chemical network are listed in Table \ref{table:grainsurfacereactions}.
In future work, the network will be extended to include photodissociation of species on the grain surface, in accordance with those networks used to model surface chemistry in interstellar and circumstellar regions.

\paragraph*{Langmuir-Hinshelwood mechanism}            \label{par:model:chem:grainsurface:lh}

In the diffusive LH mechanism, both reactants move over the surface and have some probability of reaction upon meeting.
The rate of a reaction where ice species $i$ and $j$ scan the surface and react is calculated as
\begin{equation}		\label{eq:klh}
	k_\mathrm{LH} = \mathrm{exp}\left(-\frac{E_a}{T_\mathrm{dust}}\right)\ (k_\mathrm{scan,i} + k_\mathrm{scan,j})\ \frac{1}{n_\mathrm{dust}} \ \ \ \mathrm{s}^{-1},
\end{equation}
where $n_\mathrm{dust}$ is the dust number density [cm$^{-3}$].
The first factor describes the probability that the reaction barrier is crossed during the encounter, with $E_a$ the activation energy [K], and $T_\mathrm{dust}$ the dust temperature [K].
The scanning rates $k_\mathrm{scan,i}$ and $k_\mathrm{scan,j}$ with which species $i$ and $j$ move over the surface are calculated as
\begin{equation}
	k_\mathrm{scan,i} = \frac{k_\mathrm{hop,i}}{N_s\ n_\mathrm{dust}^{-1}},
\end{equation} 
where $N_s$ is the number density of grain surface sites and $n_\mathrm{dust}$ is the dust grain number density, so that $N_s\ n_\mathrm{dust}^{-1}$ corresponds to the average total number of binding sites per grain.
The thermal hopping rate of species $i$, $k_\mathrm{hop,i}$, is determined by
\begin{equation}
	k_\mathrm{hop,i} = \nu\ \mathrm{exp}\left(-\frac{E_\mathrm{diff,i}}{T_\mathrm{dust}}\right).
\end{equation}
The diffusion barrier $E_\mathrm{diff}$ is difficult to determine experimentally. It is assumed to be a universal fixed fraction $f$ of the binding energy of the ice species, which is likely to lie around $0.3-0.4$ \citep{Karssemeijer2014}. We assume that $f = 0.4$.

\paragraph*{Eley-Rideal mechanism}            \label{par:model:chem:grainsurface:er}

In the ER mechanism, a gas-phase reactant collides with a (stationary) reactant adsorbed on the grain surface. 
The reaction rate is simpler than that of the LH mechanism, because the two reactants have only a single attempt to cross the reaction barrier, and is given by
\begin{equation}		\label{eq:ker}
	k_\mathrm{ER} = \theta\ \mathrm{exp}\left(-\frac{E_a}{T_\mathrm{dust}}\right)\ S \ \langle v \rangle \ \sigma_\mathrm{dust} \ F(Q) \ \ \ \mathrm{s}^{-1},
\end{equation}
where $\theta$ is the grain-surface coverage factor of the adsorbed species, and the exponential again describes the probability that the reaction barrier is crossed. 
The factor $S \langle v \rangle \sigma_\mathrm{dust} F(Q)$ describes the rate of arrival of the gas-phase reactant (the accretion rate, Eq. \ref{eq:kaccr}).
The effects of a sub-monolayer regime are taken into account when calculating the surface coverage factor.

\subsection{Description of the dust}                \label{subsect:model:dust}

\begin{table}
	\centering
	\caption{Parameters of the dust-grain size distribution.}
	\label{table:dustparams}
	\begin{tabular}{ll} 
		\hline
	Minimum grain size, $a_\mathrm{min}$ & $5 \times 10^{-7}$ cm \\
	Maximum grain size, $a_\mathrm{min}$ & $0.25 \times 10^{-4}$ cm \\
	Dust-to-gas mass ratio, $\psi$			& $2 \times 10^{-3}$ \\
	Surface density of binding sites, $n_s$	& $10^{15}$ cm$^{-2}$ \\
	Silicate dust bulk density$^1$, $\rho_\mathrm{dust,bulk}$ 	& 3.5 g cm$^{-3}$  \\
	Carbonaceous dust bulk density$^1$, $\rho_\mathrm{dust,bulk}$ 	& 2.24 g cm$^{-3}$  \\
		\hline
	\end{tabular}
    \footnotesize
    { {{References.}} (1) \citet{Nomura2005}
    }

\end{table}

\subsubsection{Dust-grain size distribution}        \label{subsubsect:model:dust:gsd}

Dust grains are already present within the outflow, i.e. it is assumed that they have formed in the inner wind within 20 R$_*$, at the start of the chemical model. They are assumed to follow the size distribution of \citet{Mathis1977},
\begin{equation}
    \frac{\mathrm{d}n}{\mathrm{d}a} \sim a^{-3.5},
\end{equation}
where $a$ is the radius of the assumed compact spherical grains.
The total dust mass within the outflow, $M$, is given by
\begin{equation}		\label{eq:totaldustmass}
    M = C \int_{a_\mathrm{min}}^{a_\mathrm{max}} a^{-3.5} \rho_\mathrm{dust,bulk} \left( \frac{4\pi}{3} a^3 \right) da = \psi \ \rho_\mathrm{gas} \ \ \  \mathrm{g\ cm^{-3}},
\end{equation}
where $C$ is a constant of proportionality, $\rho_\mathrm{dust,bulk}$ is the bulk density of the dust grains [g cm$^{-3}$], $\rho_\mathrm{gas}$ is the gas mass density [g cm$^{-3}$], $\psi$ is the dust-to-gas mass ratio, and $a_\mathrm{min}$ and $a_\mathrm{max}$ are the minimum and maximum grain size [cm], respectively.
Together with an assumed value of $\psi$, the constant of proportionality $C$ can be calculated.
This enables us to calculate the average dust grain cross section per unit volume,
\begin{equation}    \label{eq:sigmadust}
    \sigma_\mathrm{dust} = C \int_{a_\mathrm{min}}^{a_\mathrm{max}} a^{-3.5} \left( \pi a^2 \right) da\ \ \  \mathrm{cm^{-1}},
\end{equation}
the dust grain number density,
\begin{equation}        \label{AmCDHSeq:ndust}
    n_\mathrm{dust} = C \int_{a_\mathrm{min}}^{a_\mathrm{max}} a^{-3.5} da \ \ \ \mathrm{cm^{-3}},
\end{equation}
and the number density of dust grain surface sites,
\begin{equation}        \label{eq:NS}
    N_s = C\ n_s \int_{a_\mathrm{min}}^{a_\mathrm{max}} a^{-3.5} \left(4 \pi a^2 \right) da \ \ \ \mathrm{cm^{-3}},
\end{equation}
where $n_s$ is the density of surface sites on the grain [cm$^{-2}$].
In this paper, we assume a single dust-grain size distribution and dust-to-gas mass ratio for all models, varying only the dust bulk density between the O-rich and C-rich outflows.
Table \ref{table:dustparams} lists the assumed values of all free parameters.

Additionally, the effect of the dust-to-gas mass ratio on the extinction of interstellar UV photons is taken into account.
The extinction due to dust throughout the outflow is assumed to be equal to $1.87 \times 10^{21}$ atoms cm$^{-2}$ mag$^{-1}$ \citep{Cardelli1989}. 
This value was derived for the ISM with $\psi \sim 0.01$. 
The effect of adopting a different $\psi$ more typical of AGB outflows is included by scaling the canonical value.

\begin{table*}
    \caption{Values of the free parameters $T_\mathrm{dust,*}$ and s as obtained by fitting Eq. (\ref{eq:tdust}) to the dust temperature profiles calculated by MCMax for different outflow densities, determined by the combination of $\dot{M}$ (M$_\odot$ yr$^{-1}$) and v$_\mathrm{exp}$ (km s$^{-1}$), and dust compositions. The chi square value of each fit is listed as well.
    }
    \centering
    \begin{tabular}{c l c c c c c c c c c}
    \hline  
    & & \multicolumn{9}{c}{Oxygen-rich} \\
    \cmidrule(lr){3-11} 
    	& & \multicolumn{3}{c}{Olivine } & \multicolumn{3}{c}{Olivine} & \multicolumn{3}{c}{Melilite} \\
    	& & \multicolumn{3}{c}{with Fe} & \multicolumn{3}{c}{without Fe} & \multicolumn{3}{c}{} \\
    \cmidrule(lr){1-2}\cmidrule(lr){3-5} \cmidrule(lr){6-8} \cmidrule(lr){9-11}  
    $\dot{M}$ & $\mathrm{v_{exp}}$ & $\mathrm{T_{dust,*}}$ & $\mathrm{s}$ & $\chi^2$ & $\mathrm{T_{dust,*}}$ & $\mathrm{s}$ & $\chi^2$ & $\mathrm{T_{dust,*}}$ & $\mathrm{s}$ & $\chi^2$ \\
    \cmidrule(lr){1-2}\cmidrule(lr){3-5} \cmidrule(lr){6-8} \cmidrule(lr){9-11}  
    10$^{-5}$ & 5 	& 1000 & 1.4 & 90 & 800 & 1.6 & 34 & 550 & 1.5 & 17 \\
    10$^{-5}$ & 15	& 1000 & 1.4 & 71 & 700 & 1.7 & 19 & 500 & 1.7 & 12 \\
    10$^{-6}$ & 5 	& 1000 & 1.5 & 73 & 600 & 1.9 & 10 & 550 & 1.5 & 10 \\
    10$^{-6}$ & 15 	& 1050 & 1.4 & 67 & 600 & 1.8 & 8.4 & 550 & 1.5 & 10 \\
    10$^{-7}$ & 5 	& 1100 & 1.3 & 67 & 550 & 2.0 & 8.2 & 500 & 1.7 & 10 \\
    10$^{-7}$ & 15	& 1100 & 1.3 & 66 & 550 & 2.0 & 7.3 & 500 & 1.5 & 11 \\
   \cmidrule(lr){1-2} \cmidrule(lr){3-11} 
    & & \multicolumn{9}{c}{Carbon-rich} \\
    \cmidrule(lr){3-11} 
    	& & \multicolumn{3}{c}{Amorphous} & \multicolumn{3}{c}{Amorphous} & \multicolumn{3}{c}{SiC} \\
    	& & \multicolumn{3}{c}{carbon CDE} & \multicolumn{3}{c}{carbon DHS} & \multicolumn{3}{c}{} \\
    \cmidrule(lr){1-2}\cmidrule(lr){3-5} \cmidrule(lr){6-8} \cmidrule(lr){9-11}  
    $\dot{M}$ & $\mathrm{v_{exp}}$ & $\mathrm{T_{dust,*}}$ & $\mathrm{s}$ & $\chi^2$ & $\mathrm{T_{dust,*}}$ & $\mathrm{s}$ & $\chi^2$ & $\mathrm{T_{dust,*}}$ & $\mathrm{s}$ & $\chi^2$ \\
    \cmidrule(lr){1-2}\cmidrule(lr){3-5} \cmidrule(lr){6-8} \cmidrule(lr){9-11}  
    10$^{-5}$ & 5 	& 1850 & 0.5 & 17 & 1700 & 0.7 & 25 & 500 & 4.8 & 100 \\
    10$^{-5}$ & 15 	& 1800 & 0.6 & 6.3 & 1700 & 0.8 & 7.4 & 400 & 5.8 & 42 \\
    10$^{-6}$ & 5 	& 1800 & 0.7 & 3.4 & 1650 & 1.0 & 3.9 & 350 & 6.2 & 26 \\
    10$^{-6}$ & 15 	& 1850 & 0.8 & 0.6 & 1850 & 0.9 & 2.7 & 350 & 6.2 & 22 \\
    10$^{-7}$ & 5 	& 2000 & 0.8 & 3.3 & 1900 & 1.0 & 0.5 & 350 & 6.0 & 17 \\
    10$^{-7}$ & 15 	& 1950 & 0.9 & 8.3 & 2000 & 1.0 & 3.7 & 350 & 6.1 & 18 \\
    \hline
    \end{tabular}%
    \label{table:temperaturedust}
\end{table*}

\subsubsection{Dust temperature profiles}           \label{subsubsect:model:dust:dusttemp}

The dust temperature as a function of the radial distance from the centre of the star, $r$, can be derived from calculating the energy balance between the dust and gas.
For an optically thin outflow where the star emits as a blackbody spectrum, the dust temperature profile can be approximated by
\begin{equation}        \label{eq:tdust}
    T_\mathrm{dust}(r) = T_\mathrm{dust,*} \left( \frac{2r}{R_*} \right)^{-\frac{2}{4+s}},
\end{equation}
where $T_\mathrm{dust,*}$ is the stellar temperature and $s$ describes the wavelength dependency of the dust opacity, assuming that this can be described by $Q^A \sim \lambda^{-s}$, where $Q^A$ is the absorptive extinction efficiency of the dust \citep{Lamers1999}.

The continuum radiative transfer code, MCMax \citep{Min2009}, was used to retrieve dust temperature profiles specific to the dust composition and the outflow density (Table \ref{table:modelparams}).
We make the simplifying assumption that the dust component is composed of a single species only.
For O-rich outflows, either MgFeSiO$_4$ (olivine with iron), or Mg$_2$SiO$_4$ (olivine without iron) or Ca$_2$Mg$_{0.5}$Al$_2$Si$_{1.5}$O$_7$ (melilite) is considered. For C-rich outflows, either amorphous carbon or SiC is considered.
Eq. (\ref{eq:tdust}) is then fitted to the dust temperature profiles as calculated by MCMax, with $T_\mathrm{dust,*}$ and $s$ left as free parameters. Although $T_\mathrm{dust,*}$ corresponds to the stellar temperature, we consider it to be a free parameter to increase the goodness-of-fit.
The resulting combinations of the parameters $T_\mathrm{dust,*}$ and $s$ are listed in Table \ref{table:temperaturedust}.
They were obtained by identifying the minimum value of the chi square test when varying $T_\mathrm{dust,*}$ in intervals of 50 K and $s$ in intervals of 0.1. These are listed in Table \ref{table:temperaturedust} as well.
Fig. \ref{fig:tdustprofiles} shows the MCMax dust temperature profiles together with the fitted power laws for the highest density outflows.

The dust opacities used were calculated for particle shapes represented by a distribution of hollow spheres (DHS) with a filling factor of 0.8 \citep{Min2003}. Olivine with and without iron had $a_\mathrm{min}$ = 0.01 $\mu$m and $a_\mathrm{max}$ = 3 $\mu$m, with optical constants from \citet{Jager1994} and \citet{Jager2003}, respectively. 
For melilite, $a_\mathrm{min}$ = 0.29 $\mu$m and $a_\mathrm{max}$ = 0.31 $\mu$m, with optical constants from \citet{Mutschke1998} were used. 
The C-bearing grains had a size of 1 $\mu$m. 
For amorphous carbon, the opacities were calculated using both the DHS and the continuous distribution of ellipsoids (CDE) approximation, with optical constants from \citet{Preibisch1993} in both cases. 
The optical constants for SiC are from \citet{Pitman2008}.
We acknowledge the variation among the optical constants and their differences with our assumed dust-grain size distribution (Table \ref{table:dustparams}). 
However, these are the optical data currently incorporated in the Leuven version of MCMax, used for radiative transfer modelling of AGB outflows.

\section{Results}				\label{sect:results}

The effect of dust-gas chemistry on the gas-phase composition will depend on a combination of four main factors: 

(i) Density of the outflow. This governs the rate at which gas-phase species accrete onto the dust, i.e., the accumulation of ice mantles.

(ii) Temperature of the dust. The rate of thermal desorption of ice species is sensitive to the dust temperature. Hence, colder grains can host a more massive ice reservoir on the grain.

(iii) Initial composition of the outflow. Species with a larger binding energy are less susceptible to thermal desorption and sputtering. 

(iv) Drift velocity between the dust and the gas. Larger drift velocities lead to faster accretion of gas-phase species. However, for $v_\mathrm{drfit} > 10$ km s$^{-1}$, sputtering starts to become important, partly destroying the accreted ice mantles.

In the next sections, we present the results of our calculations and discuss the role of each of the listed factors in setting the gas-phase abundances through the outflow.

\begin{figure*}
 \includegraphics[width=\textwidth]{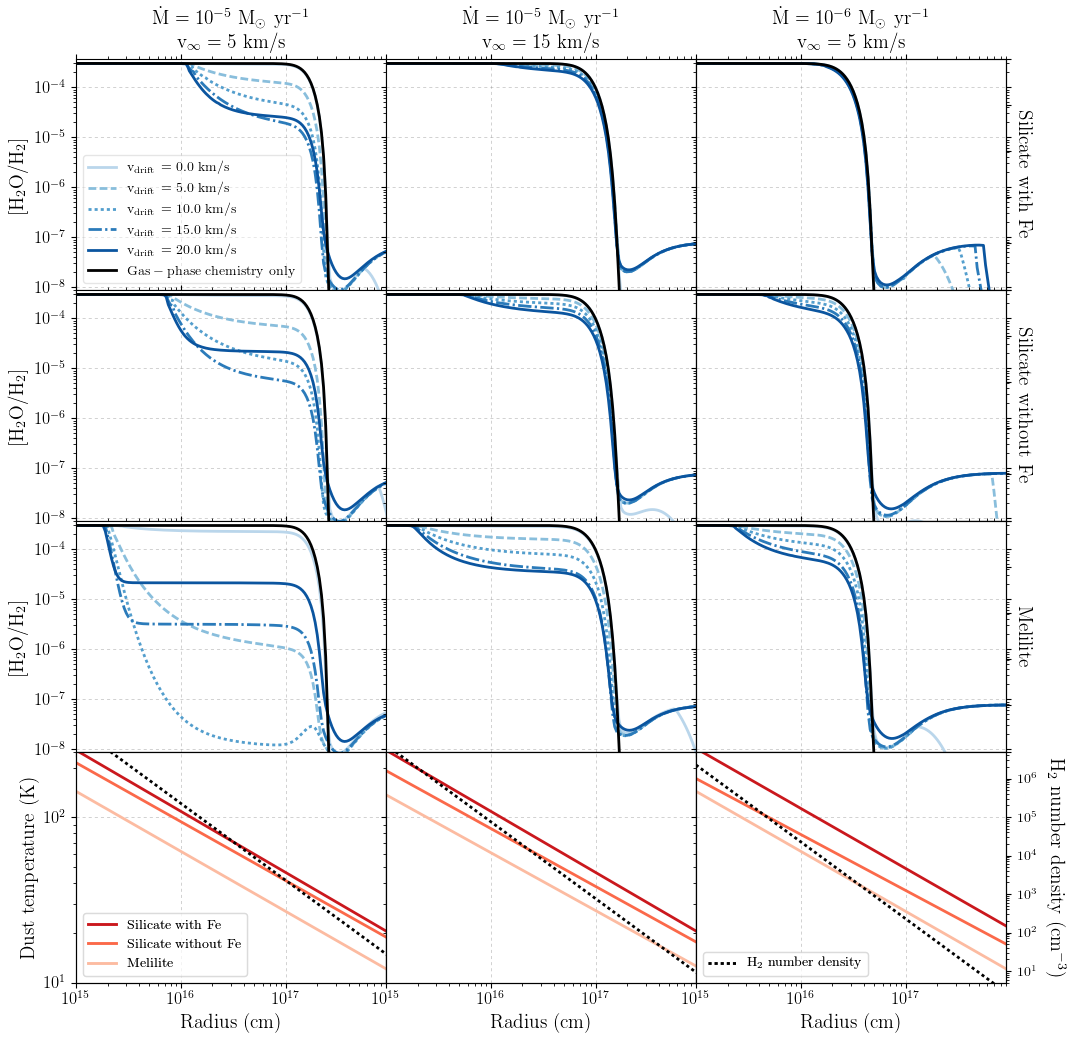}
 \caption{Upper three rows: abundance of H$_2$O with respect to H$_2$ throughout oxygen-rich outflows for different types of dust (rows) and for different density outflows (columns). Black: abundance profile obtained without including dust-gas chemistry. Colours: abundance profiles obtained when including dust-gas chemistry, where different colours and linestyles correspond to different drift velocities $v_\mathrm{drift}$. 
 Final row: H$_2$ number density (dotted black line) and the different dust temperature profiles (red solid lines) for each outflow density.
 }
 \label{fig:orich-depletion-h2o}
\end{figure*}

\begin{figure*}
 \includegraphics[width=\textwidth]{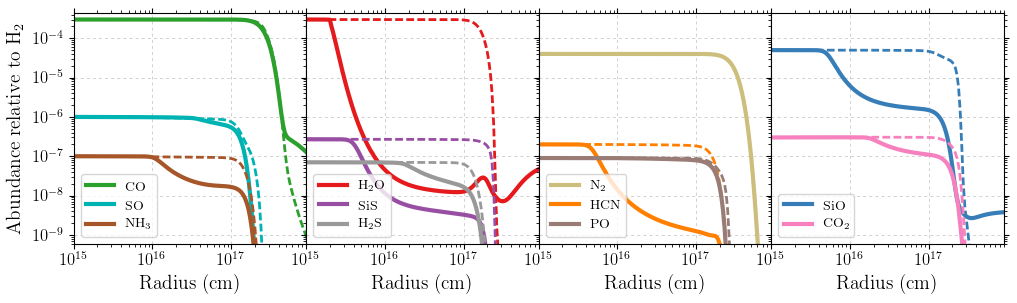}
 \caption{Abundance profiles of the O-rich parent species (Table \ref{table:Model-Parents}) without dust-gas chemistry (dashed lines) and with dust-gas chemistry (solid lines) in an outflow with $\dot{M} = 10^{-5}$ M$_\odot$ yr$^{-1}$, $v_\infty = 5$ km s$^{-1}$, and $v_\mathrm{drift} = 10$ km s$^{-1}$ and melilite dust (corresponding to the outflow with the maximum H$_2$O depletion in Fig. \ref{fig:orich-depletion-h2o}). 
 }
 \label{fig:orich-maxdepletion-parents}
\end{figure*}

\begin{figure*}
 \includegraphics[width=\textwidth]{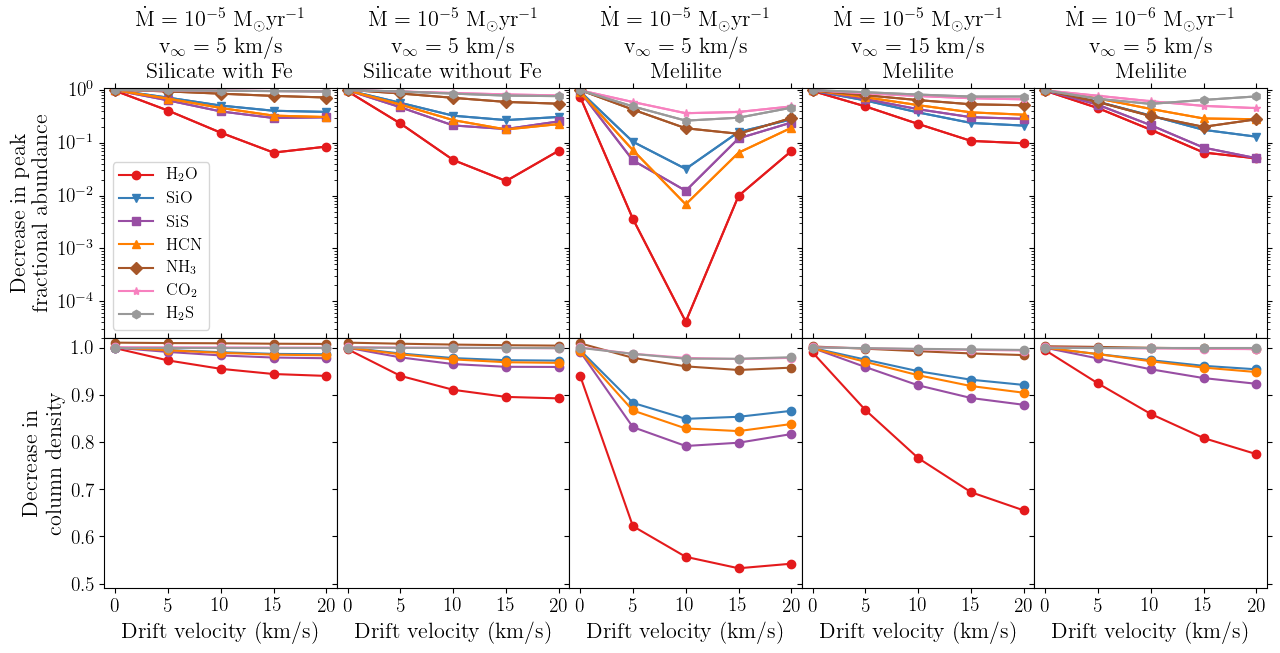}
 \caption{Schematic view of dust-gas chemistry of the depletion of parent species in O-rich outflows.
 Upper panel: maximal decrease in peak fractional abundance in the region before $r < 10^{17}$ cm for $\dot{M} = 10^{-5}$ M$_\odot$ yr $^{-1}$ and $r < 4 \times 10^{16}$ cm for $\dot{M} = 10^{-6}$ M$_\odot$ yr $^{-1}$ (corresponding to the region not affected by photodissociation).
 Lower panel: decrease in column density when including dust-gas chemistry.
 }
 \label{fig:orich-depletion-parents}
\end{figure*}

\subsection{Oxygen-rich outflows}                \label{subsect:results:orich}

Fig. \ref{fig:orich-depletion-h2o} shows the effect of different outflow densities, types of dust, and drift velocities on the H$_2$O abundance throughout an O-rich outflow. 
Also shown are the physical structures (H$_2$ number density and dust temperature as a function of radius) for each model tested (bottom row).
The H$_2$O abundance can decrease by up to four orders of magnitude: the largest effects are seen for the highest density outflow, with $\dot{M} = 10^{-5}$ M$_\odot$ yr$^{-1}$ and $v_\infty = 5$ km s$^{-1}$, and when including colder melilite dust.
For outflows less dense than an outflow with $\dot{M} = 10^{-6}$ M$_\odot$ yr$^{-1}$ and $v_\infty = 5$ km s$^{-1}$, dust-gas chemistry does not significantly influence the gas-phase water abundance.
H$_2$O is depleted onto dust grains when the dust is colder than $\sim$ 100 K, which occurs closer to the star for melilite. 
A higher drift velocity between dust and gas increases the depletion of gas-phase water onto dust, because more material is swept up by the dust.
For $v_\mathrm{drift} > 10$ km s$^{-1}$ sputtering starts to destroy the ice mantle, reversing this trend. 
However, sputtering is not able to remove the entire ice mantle, so that gas-phase H$_2$O is depleted even for the extreme case of $v_\mathrm{drift} = 20$ km s$^{-1}$.

\begin{figure*}
 \includegraphics[width=0.9\textwidth]{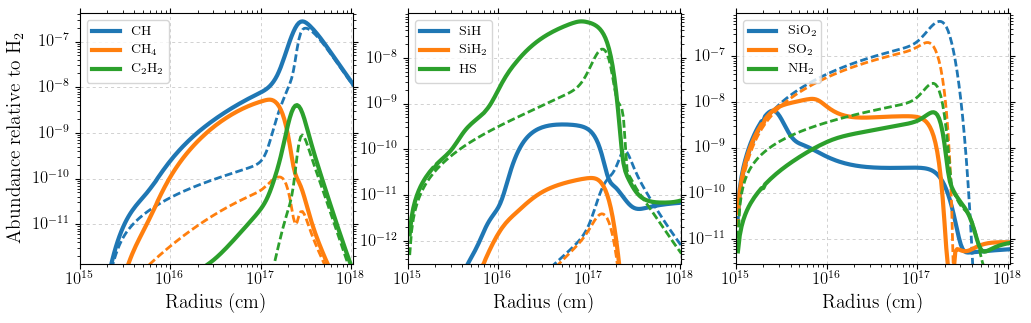}
 \caption{Abundance profiles of daughter species affected by dust-gas chemistry in an O-rich outflow with $\dot{M} = 10^{-5}$ M$_\odot$ yr$^{-1}$, $v_\infty = 5$ km s$^{-1}$, and $v_\mathrm{drift} = 10$ km s$^{-1}$ and melilite dust (corresponding to the outflow with the maximum H$_2$O depletion in Fig. \ref{fig:orich-depletion-h2o}). Solid lines: abundance profile obtained when including dust-gas chemistry. Dashed lines: abundance profiles obtained in the gas-phase only model. 
 }
 \label{fig:orich-daughters}
\end{figure*}

Fig. \ref{fig:orich-maxdepletion-parents} shows the abundance of the O-rich parent species (Table \ref{table:Model-Parents}) in the outflow with the maximum depletion of parent species onto the dust, namely melilite dust in the highest density outflow, characterised by $\dot{M} = 10^{-5}$ M$_\odot$ yr$^{-1}$, $v_\infty = 5$ km s$^{-1}$, and $v_\mathrm{drift} = 10$ km s$^{-1}$. 
The effect of the binding energy can be clearly seen in this Figure. 
CO and N$_2$, with binding energies below 1000 K, are barely affected by dust-gas interactions, while species with higher binding energies are readily depleted onto the dust.
The largest depletion is seen for H$_2$O, the species with the largest binding energy of 4880 K, for which the abundance decreases by four orders of magnitude.
The SiO, SiS, and HCN abundances decrease by up to two orders of magnitude, and those of NH$_3$, H$_2$S and CO$_2$ by up to an order of magnitude.
Fig. \ref{fig:orich-parents-appendix} shows the behaviour of the parent species in the other outflows shown in Fig. \ref{fig:orich-depletion-h2o}, with either a different dust type or density profile.

A schematic view of the degree of depletion of all parent species as a function drift velocity in the O-rich outflow is given in Fig. \ref{fig:orich-depletion-parents}. We show the results for all three dust grain types in the highest density outflow with $\dot{M} = 10^{-5}$ M$_\odot$ yr$^{-1}$ and $v_\infty = 5$ km s$^{-1}$, and for the coldest melilite dust in lower density outflows.
The decrease in peak fractional abundance (upper panel) is calculated in the region which is unaffected by photodissociation, i.e., the inner and intermediate outflow, where $r < 10^{17}$ cm for $\dot{M} = 10^{-5}$ M$_\odot$ yr$^{-1}$ and $r <  4 \times 10^{16}$ cm for $\dot{M} = 10^{-6}$ M$_\odot$ yr$^{-1}$.
The highest depletion occurs for $v_\mathrm{drift} = 10-15$ km s$^{-1}$.
The effects of dust temperature and binding energy are also clearly seen.
The column densities of the parent species (lower panel) do not significantly decrease, because depletion only affects the parent abundances from $5 \times 10^{15} - 1 \times 10^{16}$ cm onwards.
The largest decrease in column density, for H$_2$O in the highest density outflow with cold melilite, is only about a factor of two.
Hence, dust-gas chemistry does not significantly affect the column densities of the O-rich parent species, despite having a significant impact on their abundance distributions throughout the outflow.

The abundance of daughter species, i.e., those species formed via chemistry from the initial parent species, is also influenced by dust-gas chemistry through two main pathways, as shown in Fig. \ref{fig:orich-daughters}.
Gas-phase daughter species with a high binding energy are depleted onto the dust (right panel).
The peak abundance of SiO$_2$ ($E_\mathrm{bind}$ = 4300 K) decreases by some three orders of magnitude and that of SO$_2$ ($E_\mathrm{bind}$ = 3010 K) by more than one order of magnitude.
The behaviour of their abundance profiles is also substantially changed.
The NH$_2$ abundance ($E_\mathrm{bind}$ = 770 K)  decreases by about a factor of five throughout the outflow.
The gas-phase abundance of certain other daughter species increases (left and middle panels).
The peak abundances and abundance profiles of hydrates are affected by dust-gas chemistry.
They are efficiently formed on the grain surface through the Langmuir-Hinshelwood mechanism and are injected back into the gas phase through sputtering.
This increases their peak abundance and changes their behaviour throughout the outflow, as their molecular shells become wider and shift closer to the star. 
The latter effect is most clearly seen for SiH.

\begin{figure*}
 \includegraphics[width=0.73\textwidth]{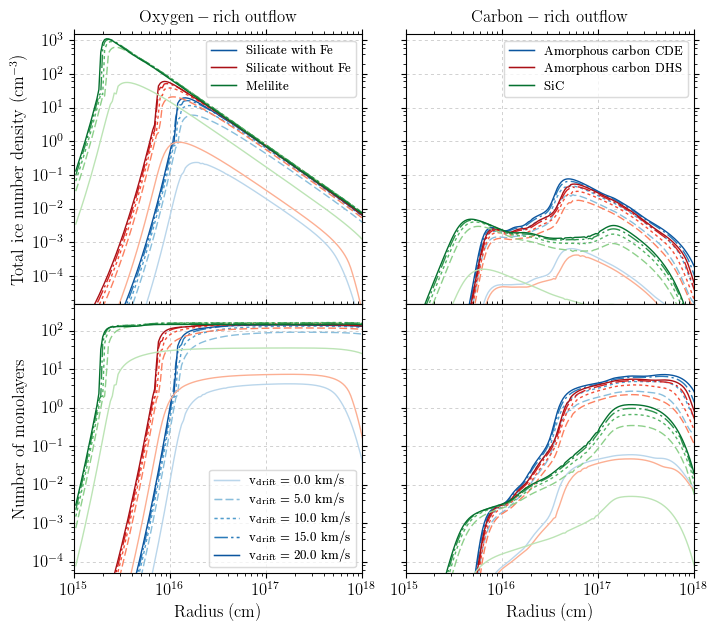}
 \caption{Total ice number density (upper panel) and number of monolayers on the dust grains (lower panel) in the highest overall density outflow with $\dot{M} = 10^{-5}$ M$_\odot$ yr$^{-1}$ and $v_\infty = 5$ km s$^{-1}$ for all dust compositions and drift velocities $v_\mathrm{drift}$. Left panels: results for the O-rich outflow. Right panels: results for the C-rich outflow.
 }
 \label{fig:totalice}
\end{figure*}

The left panels of Fig. \ref{fig:totalice} show the overall effect of the drift velocity and the dust composition on the ice formation in the highest density O-rich outflow with $\dot{M} = 10^{-5}$ M$_\odot$ yr$^{-1}$, $v_\infty = 5$ km s$^{-1}$. 
The total ice number density (upper panel) and number of monolayers (lower panel) increase as the dust temperature decreases and the drift velocity increases. 
The smallest total ice number densities, and therefore the number of monolayers, are found in outflows with zero drift velocity, i.e. where the dust and gas have the same outflow velocity, which is generally not the case in dust-driven AGB outflows.
While sputtering decreases the total ice number density for $v_\mathrm{drift} \gtrapprox 10$ km s$^{-1}$, the ice mantle is never fully destroyed by dust-gas collisions.
Colder dust grains have a larger total ice abundance.
The dust temperature profile affects the total ice abundance throughout the outflow, where melilite dust has its peak in total ice abundance closer to the star.

Hence, dust-gas interactions can significantly affect the gas-phase chemistry in O-rich outflows.
The largest effects are seen for H$_2$O, the parent species with the largest binding energy. 
Other parent species with large binding energies are also depleted onto the dust, while CO and N$_2$ experience negligible effects on their abundances.
Colder dust leads to a larger depletion of gas-phase species, as does a larger overall outflow density.
A non-zero drift velocity increases the total amount of ice significantly. 
While sputtering destroys the ice mantle for $v_\mathrm{drift} \gtrapprox 10$ km s$^{-1}$, the total ice abundance is still larger than for outflows without drift between dust and gas.
Daughter species can also be depleted onto dust, if their binding energies are sufficiently high.
An increase in abundance is also possible, especially for hydrates, as they are formed on the grain surface and released into the gas phase through sputtering. 

\subsection{Carbon-rich outflows}                \label{subsect:results:crich}

\begin{figure*}
 \includegraphics[width=\textwidth]{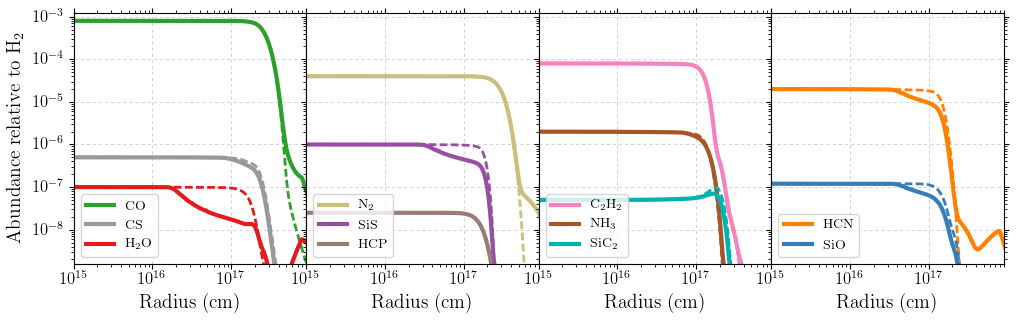}
 \caption{Abundance profiles of the C-rich parent species (Table \ref{table:Model-Parents}) without dust-gas chemistry (dashed lines) and with dust-gas chemistry (solid lines) in an outflow with $\dot{M} = 10^{-5}$ M$_\odot$ yr$^{-1}$, $v_\infty = 5$ km s$^{-1}$, and $v_\mathrm{drift} = 10$ km s$^{-1}$ and amorphous carbon CDE dust. 
 }
 \label{fig:crich-maxdepletion-parents}
\end{figure*}

Fig. \ref{fig:crich-maxdepletion-parents} shows the abundance of the C-rich parent species (Table \ref{table:Model-Parents}) in the outflow with the maximum depletion onto dust, namely amorphous CDE carbon in an outflow with $\dot{M} = 10^{-5}$ M$_\odot$ yr$^{-1}$, $v_\infty = 5$ km s$^{-1}$, and $v_\mathrm{drift} = 10$ km s$^{-1}$.
Fig. \ref{fig:crich-parents-appendix} shows the behaviour of the parent species for the other two dust grain types.
Most parent species do not deplete efficiently onto the dust because of their smaller binding energies. 
The H$_2$O abundance decreases by almost an order of magnitude, while those of SiS ($E_\mathrm{bind} = 3800$ K), SiO ($E_\mathrm{bind} = 3500$ K) and HCN ($E_\mathrm{bind} = 3610$ K) decrease by a factor of a few at most.

\begin{figure}
\centering
 \includegraphics[width=0.8\columnwidth]{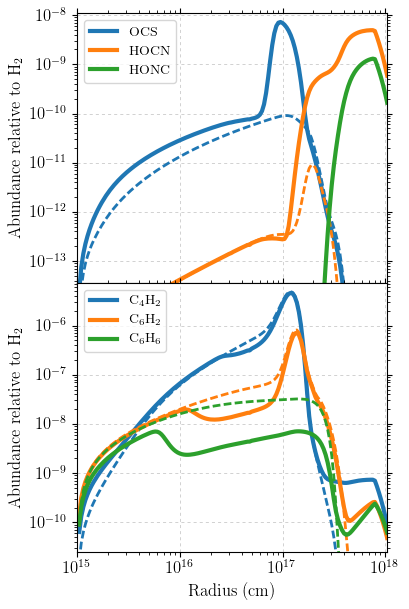}
 \caption{Abundance profiles of daughter species affected by dust-gas chemistry in an C-rich outflow with $\dot{M} = 10^{-5}$ M$_\odot$ yr$^{-1}$, $v_\infty = 5$ km s$^{-1}$, and $v_\mathrm{drift} = 10$ km s$^{-1}$ and amorphous carbon CDE dust. Solid lines: abundance profile obtained when including dust-gas chemistry. Dashed lines: abundance profiles obtained without including dust-gas chemistry. 
 }
 \label{fig:crich-daughters}
\end{figure}

Fig. \ref{fig:crich-daughters} shows the effect of dust-gas interactions on the fractional abundances of daughter species throughout the outflow, where we again find that certain daughter species are formed on the grain surface, while others are depleted onto the dust.
Through O-addition reactions, O-bearing molecules are formed on the dust and released into the gas phase through sputtering (upper panel).
This leads to sharp increases in peak abundance in the outer wind, with that of OCS increasing almost two orders of magnitude, that of HOCN more than two orders of magnitude, and that of HONC more than ten orders of magnitude.
C-chains are depleted onto the dust (lower panel). 
Since longer C-chains have higher binding energies, their abundances decrease the most: while the C$_4$H$_2$ abundance ($E_\mathrm{bind} = 4187$ K) decreases only slightly, that of C$_6$H$_6$ (benzene; $E_\mathrm{bind} = 7587$ K) decreases by roughly an order of magnitude.
The depletion of C$_4$H$_2$ in turn affects that of C$_6$H$_2$, as the gas-phase reaction between C$_4$H$_2$ and C$_2$H is one of its main formation pathways.
Note that a limited chemistry only for benzene is included in the network, involving only gas-phase reactions and dust-gas interactions.

The right panels of Fig. \ref{fig:totalice} show the total ice number density (upper panel) and number of monolayers (lower panel) in the highest density C-rich outflow with $\dot{M} = 10^{-5}$ M$_\odot$ yr$^{-1}$, $v_\infty = 5$ km s$^{-1}$ for different drift velocities and the three types of C-rich dust. 
The total ice number density and number of monolayers throughout the outflow depend on the temperature profile of the dust, with clear differences between the two amorphous carbonaceous grains and SiC dust. 
Again, the smallest total ice number densities and number of monolayers are found in outflows with zero drift velocity.
A higher drift velocity leads to larger total ice number density and number of monolayers throughout the outflow.

\subsection{Comparison between O- and C-rich outflows}                \label{subsect:results:comparison}

In contrast to the O-rich outflows, we find that the gas-phase chemistry in C-rich outflows is less influenced by dust-gas chemistry.
This is largely due to the lower binding energies of the C-rich parent species and the grains being warmer compared to those in an O-rich outflow.  This is because warmer grains lead to depletion occurring further out in the outflow, where the density and hence the accretion rate are smaller, resulting in a less massive ice reservoir.
Nonetheless, the effect on higher density C-rich outflows is not negligible, with depletion of some higher binding energy parent species (e.g., H$_2$O, SiS, and HCN) and daughter species (e.g., C$_6$H$_2$ and C$_6$H$_6$), as well as the formation of some daughter species on the grain surface (e.g., OCS and HOCN).

The smaller effect on the gas phase is reflected in the lower total ice number density and fewer number of monolayers.
The maximum total ice number density in C-rich outflows is four orders of magnitude lower than that of the O-rich outflow. 
Similarly, the maximum number of monolayers is approximately two orders of magnitude lower (Fig. \ref{fig:totalice}).
This lower dust grain coverage in C-rich outflows also leads to a different behaviour with increasing drift velocity, because the sputtering rate is much lower. 
The total ice number density increases with drift velocity, whereas in O-rich outflows, it decreases as $v_\mathrm{drift} \gtrapprox 10$ km s$^{-1}$.

\section{Discussion}					\label{sect:discussion}

Including dust-gas chemistry in the chemical kinetics model can considerably affect the gas-phase chemistry throughout the outflow.
This depends on four main factors, namely (i) the density of the outflow, (ii) the dust temperature, (iii) the initial gas-phase composition, and (iv) the drift velocity between the dust and gas (Sect. \ref{sect:results}).
A general effect is that molecules persist in the gas phase to larger radii, because ice species are photodesorbed off the dust grains and released back into the gas phase in the outermost tenuous region.

The dust grains included in our models are composed of a single species only, which is a simplified assumption. A mixture of dust species will lead to a different dust temperature profile, e.g., including silicates with or without Fe in melilite increases the dust temperature.
Additionally, the binding energy of the ice species to the grain surface plays a crucial role in setting the level of depletion. 
However, not all species have measured binding energies, in which case estimates are used. Only the binding energies of the following parents have been measured in the laboratory: CO and N$_2$ \citep{Oberg2005}; H$_2$O and \ce{NH3} \citep{Brown2007}; HCN \citep{Noble2013}; \ce{CO2} \citep{Noble2012}; \ce{H2S} \citep{Collings2004}; and \ce{C2H2} \citet{Smith2016}.
For the daughter species discussed, only CH$_4$, \ce{C2H2}, \ce{C2H6}, and \ce{SO2} have measured binding energies \citep{Collings2004,Oberg2009,Smith2016}.
For the other species, we used approximations \citep[see, e.g.,][and references therein]{Allen1977}, as listed in the \textsc{Rate12} database or in \citet{Penteado2017}.

In Sects \ref{subsect:discussion:parents} and \ref{subsect:discussion:daughters}, we discuss how the distribution of the parent species and the formation of daughter species are influenced, respectively.
Sect. \ref{subsect:discussion:sputtering} considers the influence of the description of the sputtering yield.
Our results are compared to observations in Sect. \ref{subsect:discussion:comparison}.

\subsection{Effect on parent species}                \label{subsect:discussion:parents}

Dust-gas chemistry can cause a decrease in peak fractional abundance of gas-phase species of several orders of magnitude.
The largest depletion is seen for H$_2$O in high density, O-rich outflows with cold (melilite) dust, with a decrease in the gas-phase abundance of four orders of magnitude.
Because depletion onto dust can only occur once the dust grains are sufficiently cold, the abundance at the start of the model is not affected.
Therefore, the total column density does not decrease significantly, maximally up to a factor of two for H$_2$O in the highest density, O-rich outflow with melilite dust.

In higher density outflows with cold dust, the accretion rate of gas-phase species is faster, while the thermal desorption rate is slower, leading to a higher build-up of ices. 
These effects are seen in each row of Fig. \ref{fig:orich-depletion-h2o}.
The most abundant O-rich parents species have a larger binding energy than the most abundant C-rich parent species (Table \ref{table:Model-Parents}), leading to a slower thermal desorption rate and therefore a larger total ice abundance (Fig. \ref{fig:totalice}).

A higher drift velocity between the dust and the gas leads to a larger accretion rate and also to an increase in sputtering efficiency from $v_\mathrm{drift} \gtrapprox 10$ km s$^{-1}$. 
Depletion is less significant for lower $v_\mathrm{drift}$, because the dust does not efficiently sweep up material, and for higher $v_\mathrm{drift}$, where sputtering decreases the abundance of the accreted ice mantle.
The drift velocity which causes the maximum depletion depends on both the outflow density and the temperature profile of the dust, but generally lies around 10 km s$^{-1}$. 

The grain size distribution assumed is the same for the O-rich and C-rich outflow. 
Only the dust bulk density $\rho_\mathrm{dust,bulk}$ depends on the initial composition (Table \ref{table:dustparams}). 
However, this does not cause significant differences between the two outflows.
The bulk density of carbonaceous material is lower than that of silicate material, which leads to a larger average dust grain cross section and number density of binding sites (Eqs.~\ref{eq:sigmadust} and \ref{eq:NS}).
Hence, a lower $\rho_\mathrm{dust,bulk}$ increases the rate of all dust-gas interactions. 
Despite accretion occurring faster, the accreted ice mantles are also more rapidly removed through thermal desorption, photodesorption and sputtering, leading to a decrease in the total ice abundance. 
The lower mass ice mantles in the C-rich outflows (Fig. \ref{fig:totalice}) are therefore partly due to the lower $\rho_\mathrm{dust,bulk}$. 
However, increasing the dust bulk density to that of silicate material leads to an increase in total ice number density of only a factor of two.
Therefore, the lower mass ice mantles are mainly due to the higher temperature of the dust grains and the generally lower binding energies of the parent species.

\subsection{Formation of gas-phase daughter species on the dust}                \label{subsect:discussion:daughters}

Gas-phase daughter species can be affected in two ways: they can be either depleted onto the dust or increase in abundance thanks to formation on the surface through grain-surface chemistry, followed by either thermal desorption, photodesorption or sputtering (Figs \ref{fig:orich-daughters} and \ref{fig:crich-daughters}).

Depletion can be caused by accretion in the case for species with a high binding energy (such as SiO$_2$ and C$_6$H$_6$ in the O- and C-rich outflow, respectively), or by the depletion of a parent species (as is the case for C$_4$H$_2$ in the C-rich outflow).
Species such as hydrates are efficiently formed on the grain surface thanks to the high mobility of H and are subsequently released into the gas phase through thermal desorption (e.g., SiH in the O-rich outflow). 
Other mobile atoms on the surface, such as O, also lead to the increase in abundance of more complex species (e.g., OCS and HOCN in C-rich outflows). 
These mobile atoms are mainly formed through accretion, with contributions of grain-surface reactions.
The newly-formed ice species are then released into the gas phase through sputtering.
Grain-surface reactions occur mainly through the LH mechanism, with only minor contributions from the ER mechanism in the outermost region. 
In this region, the surface mobility of the ice species has decreased due to the lower dust temperature, decreasing the importance of the diffusive LH mechanism \citep{Ruffle2001}.
Additionally, photodissociation of molecules increases the abundance of gas-phase atoms. 
However, because of the lower density in this region, the ER mechanism contributes only up to a few percent to the formation of certain ice species.

Note that these results are obtained using the basic chemical network described in Sect.~\ref{sect:model}, which does not include photodissociation of ice species on the grain surface. 
In future work, we will include a more complete grain surface chemical network, allowing us to study the ice mantle composition in greater detail, as well as the return to the gas phase.

\subsection{Comparison of sputtering descriptions}                \label{subsect:discussion:sputtering}

\begin{figure}
 \includegraphics[width=0.9\columnwidth]{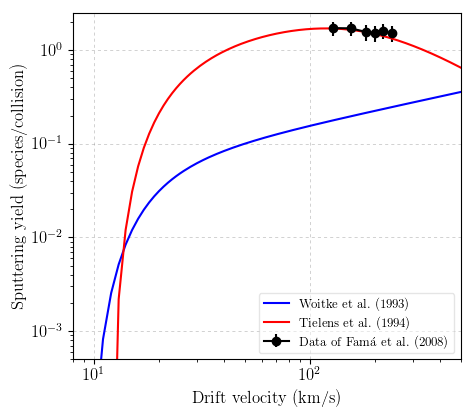}
 \caption{Sputtering yield of H$_2$O ice by He$^+$ collisions, as predicted by \citet{Woitke1993} (blue line), \citet{Tielens1994} (red line) and measured by \citet{Fama2008} (black data points).
  }
 \label{fig:sputtering-comparison}
\end{figure}

A different description of the total sputtering yield is given by \citet{Woitke1993}, who use 
\begin{equation}		\label{eq:yieldwoitke}
	Y(E) = 0.0064\ m_t\ \gamma^{5/3} \left(\frac{E}{E_{th}}\right)^{1/4} \left(1 - \frac{E_{th}}{E}\right)^{7/2},
\end{equation}
with $\gamma = 4\ m_t\ m_p/ (m_t + m_p)^2$, and $m_t$ and $m_p$ the mass of the target ice species and projectile gas-phase species [amu], respectively.
In this paper, we have adopted the description of \citet{Tielens1994}, because they use a more comprehensive description of yield measurements across a larger energy range. 
Fig. \ref{fig:sputtering-comparison} shows the sputtering yield of H$_2$O ice by He$^+$ as described by \citet{Tielens1994} (Eq. \ref{eq:ksput}) and \citet{Woitke1993} together with the experimental data of \citet{Fama2008}. 
Unlike the \citet{Woitke1993} description, that of \citet{Tielens1994} is able to reproduce the higher energy experimental data.

\begin{figure}
 \includegraphics[width=0.9\columnwidth]{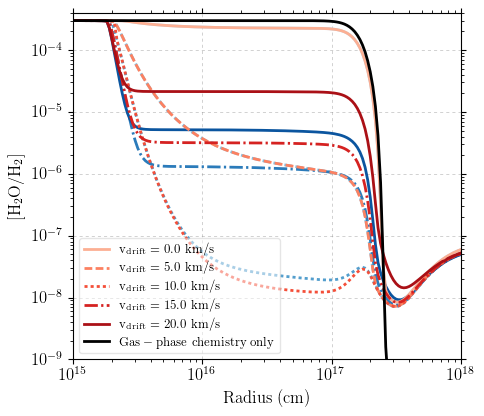}
 \caption{The effect of the different sputtering descriptions on the H$_2$O abundance profiles for different drift velocities. Red: results using the \citet{Tielens1994} description. Blue: results using the \citet{Woitke1993} description. 
 Black: abundance profile obtained without including dust-gas chemistry.
 The outflow shown is the O-rich outflow with $\dot{M} = 10^{-5}$ M$_\odot$ yr$^{-1}$, $v_\infty = 5$ km s$^{-1}$, and $v_\mathrm{drift} = 10$ km s$^{-1}$ and melilite dust (corresponding to the outflow with the maximum H$_2$O depletion in Fig. \ref{fig:orich-depletion-h2o}).
  }
 \label{fig:sputtering}
\end{figure}

Fig. \ref{fig:sputtering} shows the H$_2$O abundance in the highest density O-rich outflow with the coldest, melilite dust for different drift velocities using both descriptions.
Differences in H$_2$O depletion between the two descriptions are only noticeable for $v_\mathrm{drift} > 10$ km s$^{-1}$, when sputtering becomes efficient.
The difference is smallest for $v_\mathrm{drift} = 10$ km s$^{-1}$ and increases with drift velocity, which is directly linked to the difference in sputtering yields.

\citet{Dijkstra2003} have used the \citet{Woitke1993} description in their simple chemical model involving only accretion, thermal desorption, and sputtering of H$_2$O. 
Together with their limited chemical network, which does not include photodesorption of water ice, their use of the less comprehensive sputtering description might overestimate the depletion of H$_2$O on dust grains.
However, it is difficult to compare our results to theirs, because they do not show abundance profiles.

\subsection{Comparison to observations}                \label{subsect:discussion:comparison}

Depletion onto dust grains has often been suggested as the mechanism underlying a decrease in abundance close to the star for some gas-phase species in both O-rich and C-rich AGB outflows.
However, to date, this has not been tested using a comprehensive chemical model that includes dust-gas interactions.
In the following, we compare the results of our grid of models to observations of the outer envelope. 
We note that direct comparisons to literature data are difficult, because often either the dust-to-gas mass ratio, the drift velocity, or both, are not known nor listed in the literature.

\subsubsection{O-rich outflows}         \label{subsubsect:discussion:comparison:orich}

The depletion of SiO has been observed in several O-rich outflows. 
\citet{Bujarrabal1989} and \citet{Sahai1993} report SiO depletion after the dust formation region ($\sim 1-2 \times 10^{15}$ cm) in O-rich outflows with different mass-loss rates.
\citet{GonzalezDelgado2003} found that the SiO abundance decreases with increasing mass-loss rate, pointing towards increased adsorption onto dust in higher density outflows. 
When comparing different density outflows, our models find similar decreases in SiO abundance.

\citet{Decin2010} measured depletion of SiO in the outflow of IK Tau around 100 R$_* \approx 5 \times 10^{15}$ cm, which is characterised by $\dot{M} = 8 \times 10^{-6}$ M$_\odot$ yr$^{-1}$, $v_\infty = 17.7$ km s$^{-1}$ and $v_\mathrm{drift} = 4$ km s$^{-1}$. 
The SiO abundance decreases approximately two orders of magnitude.
Our models can reproduce this level of depletion in the outflow shown in Fig. \ref{fig:orich-maxdepletion-parents}. 
However, in an oxygen-rich outflow with melilite dust and $\dot{M} = 1 \times 10^{-6}$ M$_\odot$ yr$^{-1}$ and $v_\infty = 5$ km s$^{-1}$ (Fig. \ref{fig:orich-parents-appendix}), which corresponds to a similar $\dot{M}/v_\infty$ ratio, the depletion is limited to a fraction of an order of magnitude.
This could be due to the dust composition in IK Tau's outflow, which is thought to be (a combination of) iron-free silicate dust and corundum \citep{Gobrecht2016,Decin2017} because colder dust gives rise to a larger gas-phase depletion.

\citet{Verbena2019} found a strong coupling between SiO depletion and gas acceleration in IK Tau and WX Psc, which has $\dot{M} = 1 \times 10^{-5}$ M$_\odot$ yr$^{-1}$ and $v_\infty = 20$ km s$^{-1}$. 
The SiO abundance decreases closer to the star in the higher density outflow of WX Psc, as predicted by our models. They also predict a depletion of around half an order of magnitude, similar to our predictions for an outflow with $\dot{M} = 1 \times 10^{-5}$ M$_\odot$ yr$^{-1}$ and  $v_\infty = 15$ km s$^{-1}$.
For IK Tau, they find a similar depletion of SiO as \citet{Decin2010}, albeit closer to the star at $\sim 50$ R$_*$.

\citet{Khouri2014b} do not find any evidence of SiO depletion in the region between $10-100$ R$_*$ $\approx 5 \times 10^{14} - 10^{15}$ cm  in the outflow of W Hya, which has $\dot{M} = 1.5 \times 10^{-7}$ M$_\odot$ yr$^{-1}$ and $v_\infty = 7.5$ km s$^{-1}$ \citep{Khouri2014a}. 
\citet{VandeSandeRDor} cannot unambiguously determine whether the decline in SiO abundance around 60 R$_*$ $\approx 3 \times 10^{15}$ cm  is due to depletion onto dust or photodissociation in the low mass-loss rate R Dor, which has $\dot{M} = 1.25 \times 10^{-7}$ M$_\odot$ yr$^{-1}$, $v_\infty = 5.5$ km s$^{-1}$ and $v_\mathrm{drift} = 8.4$ km s$^{-1}$. 
For such low density outflows, our chemical models do not predict any depletion, pointing towards the observed decline in abundance as being due to photodissociation.

\citet{Decin2010} have also measured depletion of SiS in the outflow of IK Tau around 100 R$_* \approx 5 \times 10^{15}$ cm.
The SiS abundances decreases approximately four orders of magnitude some 100 R$_*$ closer to the star compared to SiO, a decline two orders of magnitude larger than that observed for SiO.
Like for SiO, our models predict a depletion limited to a fraction of an order of magnitude, which could again be due to the dust composition.
However, \citet{Danilovich2019} do not find evidence of depletion of SiS in IK Tau. 
Rather, they find a smaller distribution of SiS than that assumed by earlier models, which took the photodissociation rate of SiS to be equal to that of SiO. 
Their result points towards a decrease in abundance due to photodissociation rather than dust-gas interactions.
They also do not detect any SiS depletion in W Hya and R Dor. 
Our models do not predict any significant SiS depletion for such lower density outflows, consistent with the observations of \citet{Danilovich2019}.

Water ice has been detected in the spectra of several OH/IR stars \citep{Omont1990,Justtanont1992}.
These stars are characterised by a high mass-loss rate, generally larger than $1 \times 10^{-5}$ M$_\odot$ yr$^{-1}$.
\citet{Sylvester1999} retrieved water ice column densities between $\sim 10 - 120 \times 10^{16}$ cm$^{-2}$. 
For the highest density O-rich outflow, accounting for the different dust grain temperature profiles and drift velocities, we find column densities between $0.9 - 300 \times 10^{16}$ cm$^{-2}$, which corresponds well with the observations.
\citet{Lombaert2013} suggest a H$_2$O depletion of 50\%, which again coincides with the values predicted by our models (Fig. \ref{fig:orich-depletion-h2o}).

\begin{figure*}
 \includegraphics[width=\textwidth]{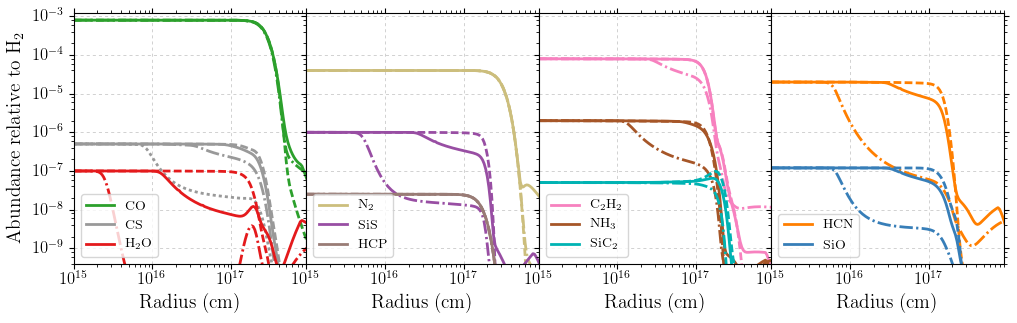}
 \caption{The effect of different dust grain temperatures on the abundance profiles of the C-rich parent species (Table \ref{table:Model-Parents}). Dashed lines: results without dust-gas chemistry. Solid lines: with dust-gas chemistry in an outflow with $\dot{M} = 10^{-5}$ M$_\odot$ yr$^{-1}$, $v_\infty = 5$ km s$^{-1}$, and $v_\mathrm{drift} = 10$ km s$^{-1}$ and amorphous carbon CDE dust (as in Fig. \ref{fig:crich-maxdepletion-parents}). Dashed-dotted lines: with dust gas chemistry and cooler melilite dust. Dotted line: with dust-gas chemistry, cooler melilte dust and binding energy of CS changed to 3200 K instead of 1900 K.
 }
 \label{fig:crich-parents-comparison}
\end{figure*}

\subsubsection{C-rich outflows}         \label{subsubsect:discussion:comparison:crich}

\citet{Schoier2006} found that the SiO abundance in C-rich outflows decreases with increasing mass-loss rate, hinting towards depletion onto dust grains.
Our highest density models show a decrease in SiO of up to half an order of magnitude at $3 \times 10^{16}$ cm.
In lower density models, SiO is not depleted onto dust grains. 
This decrease does not reproduce the depletion observed by \citet{Schoier2006}, who find a difference in SiO abundance of roughly two orders of magnitude when comparing similar outflow densities.

\citet{Agundez2012} measured that CS and SiS have significantly lower abundances in the outer region of the outflow of IRC+10216, which is characterised by $\dot{M} = 1.5 \times 10^{-5}$ M$_\odot$ yr$^{-1}$ and $v_\infty = 14.5$ km s$^{-1}$ \citep{DeBeck2010}.
The decrease in abundance occurs around $2 \times 10^{14}$ cm, so before the start of our model. 
For our models with $\dot{M} = 1.5 \times 10^{-5}$ M$_\odot$ yr$^{-1}$ and $v_\infty = 15$ km s$^{-1}$, we find a depletion of a factor of two, similar to \citet{Agundez2012}, albeit further out at $\sim 2 \times 10^{16}$ cm. 
CS is not depleted in our model.

When studying a sample of C-rich outflows, \citet{Massalkhi2019} found that the abundance of CS and SiO, as well as SiS (tentatively) decreases with increasing mass-loss rate. 
The CS abundances decreases by about a factor of five when increasing the outflow density by an order of magnitude, the SiO abundance decreases by some two orders of magnitude, while the SiS abundances tentatively decreases by a factor of five.
When comparing different outflow densities, we do not reproduce the observed depletion of any of the three molecules.

The discrepancies with observations for the C-rich outflows could be due to the temperature of the dust grains. 
Colder grains would yield larger depletion of parent species closer to the star.
Indeed, when using the dust temperature as derived for melilite dust, we find similar decreases in SiO and SiS abundances to \citet{Massalkhi2019} when comparing different density outflows.
However, this does not result in a significant CS depletion. When adjusting its binding energy from 1900 K to 3200 K, as calculated by \citet{Wakelam2017}, we do find that CS is depleted onto the dust. 
This is illustrated in Fig. \ref{fig:crich-parents-comparison} and indicates that the binding energy estimated by \citet{Garrod2006}, may be too low.
This shows the importance of accurate binding energies in chemical models and demonstrates how the comparison with observations can highlight potentially erroneous data.

\citet{Keady1993} found that NH$_3$ and SiH$_4$ are produced around $10 - 40$ R$_*$ $\approx 5 \times 10^{14} - 2 \times 10^{15}$ cm in IRC+10216.
Our models start at $10^{15}$ cm, since we assume that the dust has already formed with the grain size distribution of \citet{Mathis1977}.
Moreover, we have included NH$_3$ as a parent species, as it is thought to be formed in the inner wind \citep{Omont1988}.
However, when pushing the inner radius of the model closer to the star at $2 \times 10^{14}$ cm and removing NH$_3$ as a parent species, we find that we do not form neither NH$_3$ nor SiH$_4$ on the dust grains.
\citet{Willacy2004} proposes H$_2$O formation on dust grains in C-rich outflows to help explain its large abundance in the inner wind.
When removing H$_2$O from the list of parent species, we find that it is not formed with a  larger abundance compared to a gas-phase only model in our model with $R_\mathrm{inner} = 2 \times 10^{14}$ cm. 
Non-TE chemistry or photon-driven chemistry in the inner region, caused by an inhomogeneous outflow and/or the presence of stellar UV photons, appear to be more likely mechanisms behind H$_2$O formation in the inner CSE \citep{Cherchneff2006,Agundez2010,VandeSande2018,VandeSande2019}.

\section{Conclusions}				\label{sect:conclusion}

We presented the results of our chemical kinetics models of O-rich and C-rich AGB outflows, which are the first to include a comprehensive gas-dust chemical network. Using these models, we investigated the effects of dust-gas chemistry on the gas-phase composition, and explored the impact of the drift velocity between the dust and gas, and different dust-grain compositions. We compared our results with observations and demonstrated that the inclusion of dust-gas chemistry can explain the depletion of several species in O-rich outflows.

The influence of dust-gas chemistry depends on four main factors: (i) the overall density of the outflow, (ii) the dust temperature, (iii) the initial composition of the outflow, and (iv) the dust drift velocity.
Ice mantles are more efficiently built up in higher density outflows with colder dust grains, because accretion is proportional to the gas-phase density and the thermal desorption rate decreases with dust temperature.
Species with larger binding energies have smaller thermal desorption rates, so that O-rich outflows show larger effects.
A drift velocity between dust and gas leads to a larger accretion rate, but values from $v_\mathrm{drift} \gtrapprox 10$ km s$^{-1}$ also initiates a destruction of the ice through sputtering.

Our models agree with observations of SiO, SiS and H$_2$O depletion in O-rich outflows.
For C-rich outflows, we are not able to reproduce the observed depletion of CS, SiO and SiS. 
When using the dust temperature profile of melilite, which is much colder than that of amorphous carbon, we find a decrease in the SiO and SiS abundance similar to those observed.
Only when adjusting the binding energy of CS to a higher value, we are able to achieve a similar result for CS, emphasising the importance of accurate binding energies in chemical models.

Dust-gas chemistry in the intermediate wind can significantly influence the gas-phase chemistry and should not be ignored in chemical models of AGB outflows.
Our results show that in order to accurately interpret high-resolution observations, it is necessary to include them in chemical kinetics models.
In future work, we will expand our chemical model to investigate in detail the composition of the ice mantles formed around the dust, how they are affected by an inhomogeneous density distribution, and to explore the effect of different dust-grain size distributions on the composition of the gas as well as the ice mantles.

\section*{Acknowledgements}

MVdS acknowledges support from the Research Foundation Flanders (FWO) through grant 12X6419N.
CW acknowledges financial support from the University of Leeds and the Science and Technology Facilities Council of the United Kingdom under grant number ST/R000549/1.
TPM and LD acknowledge support from the European Research Council (ERC) under the European
Unions Horizon 2020 research and innovation programme (grant agreement No 646758, AEROSOL).
\begin{appendix}

\section{Grain-surface reactions included in the network}				\label{sect:app:grainsurfacereactions}

Table \ref{table:grainsurfacereactions} lists all grain-surface reactions included in our chemical network, together with their assumed activation energies $E_\mathrm{a}$.
Each reaction is included following both the Langmuir-Hinshelwood and Eley-Rideal mechanism.

\begin{table*}
    \caption{Grain-surface reactions included in the chemical network. Each reaction is included following both the Langmuir-Hinshelwood and Eley-Rideal mechanism. The assumed activation barrier of the reaction $E_\mathrm{a}$ (K) is also listed.
    }
    \resizebox{0.9\textwidth}{!}{%
    \centering
    \begin{tabular}{l l l l l l l l l l l l l l l l l l l}
    \hline  
    \multicolumn{3}{c}{Reactants} & & \multicolumn{3}{c}{Products} & $E_\mathrm{a}$ [K] & & \multicolumn{3}{c}{Reactants} & & \multicolumn{3}{c}{Products} & $E_\mathrm{a}$ [K]\\
    \cmidrule(lr){1-8} \cmidrule(lr){10-17}
H & + & C  & $\rightarrow$ & CH &  &   &  0.0 & & H & + & C$_5$  & $\rightarrow$ & C$_5$H &  &   &  0.0 \\
H & + & CH  & $\rightarrow$ & CH$_2$ &  &   &  0.0 & & H & + & C$_5$H  & $\rightarrow$ & C$_5$H$_2$ &  &   &  0.0 \\
H & + & CH$_2$  & $\rightarrow$ & CH$_3$ &  &   &  0.0 & & H & + & C$_6$  & $\rightarrow$ & C$_6$H &  &   &  0.0 \\
H & + & CH$_3$  & $\rightarrow$ & CH$_4$ &  &   &  0.0 & & H & + & C$_6$H  & $\rightarrow$ & C$_6$H$_2$ &  &   &  0.0 \\
H & + & CH$_4$  & $\rightarrow$ & CH$_3$ & + & H$_2$  &  5940.0 & & H & + & C$_5$N  & $\rightarrow$ & HC$_5$N &  &   &  0.0 \\
H & + & N  & $\rightarrow$ & NH &  &   &  0.0 & & H & + & C$_7$  & $\rightarrow$ & C$_7$H &  &   &  0.0 \\ 
H & + & NH  & $\rightarrow$ & NH$_2$ &  &   &  0.0 & & H & + & C$_7$H  & $\rightarrow$ & C$_7$H$_2$ &  &   &  0.0 \\
H & + & NH$_2$  & $\rightarrow$ & NH$_3$ &  &   &  0.0 & & H & + & C$_8$  & $\rightarrow$ & C$_8$H &  &   &  0.0 \\
H & + & O  & $\rightarrow$ & OH &  &   &  0.0 & & H & + & C$_8$H  & $\rightarrow$ & C$_8$H$_2$ &  &   &  0.0 \\
H & + & OH  & $\rightarrow$ & H$_2$O &  &   &  0.0 & & H & + & C$_7$N  & $\rightarrow$ & HC$_7$N &  &   &  0.0 \\
H & + & F  & $\rightarrow$ & HF &  &   &  0.0 & & H & + & C$_9$  & $\rightarrow$ & C$_9$H &  &   &  0.0 \\
H & + & C$_2$  & $\rightarrow$ & C$_2$H &  &   &  0.0 & & H & + & C$_9$H  & $\rightarrow$ & C$_9$H$_2$ &  &   &  0.0 \\
H & + & C$_2$H  & $\rightarrow$ & C$_2$H$_2$ &  &   &  0.0 & & H & + & C$_{10}$  & $\rightarrow$ & C$_{10}$H &  &   &  0.0 \\
H & + & C$_2$H$_2$  & $\rightarrow$ & C$_2$H$_3$ &  &   &  1210.0 & & H & + & C$_{10}$H  & $\rightarrow$ & C$_{10}$H$_2$ &  &   &  0.0 \\
H & + & C$_2$H$_3$  & $\rightarrow$ & C$_2$H$_4$ &  &   &  0.0 & & H & + & C$_9$N  & $\rightarrow$ & HC$_9$N &  &   &  0.0 \\
H & + & C$_2$H$_4$  & $\rightarrow$ & C$_2$H$_5$ &  &   &  750.0 & & H & + & HCO  & $\rightarrow$ & CO & + & H$_2$  &  1850.0 \\
H & + & C$_2$H$_5$  & $\rightarrow$ & CH$_3$CH$_3$ &  &   &  0.0 & & H & + & H  & $\rightarrow$ & H$_2$ &  &   &  0.0 \\
H & + & CH$_3$CH$_3$  & $\rightarrow$ & C$_2$H$_5$ & + & H$_2$  &  4890.0 & & H$_2$ & + & O  & $\rightarrow$ & OH & + & H  &  3170.0 \\
H & + & CN  & $\rightarrow$ & HCN &  &   &  0.0 & & H$_2$ & + & C  & $\rightarrow$ & CH$_2$ &  &   &  12000.0 \\
H & + & HCN  & $\rightarrow$ & H$_2$CN &  &   &  13000.0 & & H$_2$ & + & CH$_2$  & $\rightarrow$ & CH$_3$ & + & H  &  3530.0 \\
H & + & H$_2$CN  & $\rightarrow$ & CH$_2$NH &  &   &  0.0 & & H$_2$ & + & CH$_3$  & $\rightarrow$ & CH$_4$ & + & H  &  6440.0 \\
H & + & CO  & $\rightarrow$ & HCO &  &   &  2500.0 & & H$_2$ & + & NH$_2$  & $\rightarrow$ & NH$_3$ & + & H  &  6300.0 \\
H & + & HCO  & $\rightarrow$ & H$_2$CO &  &   &  0.0 & & H$_2$ & + & OH  & $\rightarrow$ & H$_2$O & + & H  &  2100.0 \\
H & + & H$_2$CO  & $\rightarrow$ & HCO & + & H$_2$  &  2500.0 & & H$_2$ & + & O$_2$H  & $\rightarrow$ & H$_2$O$_2$ & + & H  &  5000.0 \\
H & + & Si  & $\rightarrow$ & SiH &  &   &  0.0 & & H$_2$ & + & C$_2$  & $\rightarrow$ & C$_2$H & + & H  &  2100.0 \\
H & + & SiH  & $\rightarrow$ & SiH$_2$ &  &   &  0.0 & & H$_2$ & + & C$_2$H  & $\rightarrow$ & C$_2$H$_2$ & + & H  &  2100.0 \\
H & + & SiH$_2$  & $\rightarrow$ & SiH$_3$ &  &   &  0.0 & & H$_2$ & + & CN  & $\rightarrow$ & HCN & + & H  &  2070.0 \\
H & + & SiH$_3$  & $\rightarrow$ & SiH$_4$ &  &   &  0.0 & & H$_2$ & + & C$_3$  & $\rightarrow$ & C$_3$H & + & H  &  2100.0 \\
H & + & NO  & $\rightarrow$ & HNO &  &   &  0.0 & & H$_2$ & + & C$_3$H  & $\rightarrow$ & H$_2$CCC & + & H  &  2100.0 \\
H & + & HNO  & $\rightarrow$ & NO & + & H$_2$  &  750.0 & & H$_2$ & + & C$_3$H  & $\rightarrow$ & C$_3$H$_2$ & + & H  &  2100.0 \\
H & + & O$_2$  & $\rightarrow$ & O$_2$H &  &   &  300.0 & & H$_2$ & + & C$_4$  & $\rightarrow$ & C$_4$H & + & H  &  2100.0 \\
H & + & O$_2$H  & $\rightarrow$ & O & + & H$_2$O  &  0.0 & & H$_2$ & + & C$_4$H  & $\rightarrow$ & HC$_4$H & + & H  &  2100.0 \\
H & + & O$_2$H  & $\rightarrow$ & O$_2$ & + & H$_2$  &  0.0 & & H$_2$ & + & C$_5$  & $\rightarrow$ & C$_5$H & + & H  &  2100.0 \\
H & + & O$_2$H  & $\rightarrow$ & OH & + & OH  &  0.0 & & H$_2$ & + & C$_5$H  & $\rightarrow$ & C$_5$H$_2$ & + & H  &  2100.0 \\
H & + & H$_2$O$_2$  & $\rightarrow$ & H$_2$O & + & OH  &  800.0 & & H$_2$ & + & C$_6$  & $\rightarrow$ & C$_6$H & + & H  &  2100.0 \\ 
H & + & S  & $\rightarrow$ & HS &  &   &  0.0 & & H$_2$ & + & C$_6$H  & $\rightarrow$ & C$_6$H$_2$ & + & H  &  2100.0 \\
H & + & HS  & $\rightarrow$ & H$_2$S &  &   &  0.0 & & H$_2$ & + & C$_7$  & $\rightarrow$ & C$_7$H & + & H  &  2100.0 \\
H & + & H$_2$S  & $\rightarrow$ & HS & + & H$_2$  &  860.0 & & H$_2$ & + & C$_7$H  & $\rightarrow$ & C$_7$H$_2$ & + & H  &  2100.0 \\
H & + & Cl  & $\rightarrow$ & HCl &  &   &  0.0 & & H$_2$ & + & C$_8$  & $\rightarrow$ & C$_8$H & + & H  &  2100.0 \\
H & + & C$_3$  & $\rightarrow$ & C$_3$H &  &   &  0.0 & & H$_2$ & + & C$_8$H  & $\rightarrow$ & C$_8$H$_2$ & + & H  &  2100.0 \\
H & + & C$_3$H  & $\rightarrow$ & C$_3$H$_2$ &  &   &  0.0 & & H$_2$ & + & C$_9$  & $\rightarrow$ & C$_9$H & + & H  &  2100.0 \\
H & + & C$_3$H$_2$  & $\rightarrow$ & CH$_2$CCH &  &   &  1210.0 & & H$_2$ & + & C$_9$H  & $\rightarrow$ & C$_9$H$_2$ & + & H  &  2100.0 \\
H & + & H$_2$CCC  & $\rightarrow$ & CH$_2$CCH &  &   &  1210.0 & & H$_2$ & + & C$_{10}$  & $\rightarrow$ & C$_{10}$H & + & H  &  2100.0 \\
H & + & CH$_2$CCH  & $\rightarrow$ & CH$_2$CCH$_2$ &  &   &  0.0 & & H$_2$ & + & C$_{10}$H  & $\rightarrow$ & C$_{10}$H$_2$ & + & H  &  2100.0 \\
H & + & CH$_2$CCH  & $\rightarrow$ & CH$_3$CCH &  &   &  0.0 & & C & + & C  & $\rightarrow$ & C$_2$ &  &   &  0.0 \\
H & + & CH$_2$CN  & $\rightarrow$ & CH$_3$CN &  &   &  0.0 & & C & + & CH  & $\rightarrow$ & C$_2$H &  &   &  0.0 \\
H & + & CNO  & $\rightarrow$ & HCNO &  &   &  0.0 & & C & + & CH$_2$  & $\rightarrow$ & C$_2$H$_2$ &  &   &  0.0 \\
H & + & CNO  & $\rightarrow$ & HONC &  &   &  0.0 & & C & + & N  & $\rightarrow$ & CN &  &   &  0.0 \\
H & + & OCN  & $\rightarrow$ & HOCN &  &   &  0.0 & & C & + & NO  & $\rightarrow$ & CN & + & O  &  0.0 \\
H & + & OCN  & $\rightarrow$ & HNCO &  &   &  0.0 & & C & + & NO  & $\rightarrow$ & CNO &  &   &  0.0 \\
H & + & CS  & $\rightarrow$ & HCS &  &   &  1000.0 & & C & + & NS  & $\rightarrow$ & CN & + & S  &  0.0 \\
H & + & CH$_3$CHO  & $\rightarrow$ & HCO & + & CH$_4$  &  2400.0 & & C & + & OCN  & $\rightarrow$ & CO & + & CN  &  0.0 \\
H & + & CH$_3$CHO  & $\rightarrow$ & H$_2$CO & + & CH$_3$  &  2400.0 & & C & + & CNO  & $\rightarrow$ & CO & + & CN  &  0.0 \\
H & + & HCS  & $\rightarrow$ & H$_2$CS &  &   &  0.0 & & C & + & NH  & $\rightarrow$ & HNC &  &   &  0.0 \\
H & + & HCOOH  & $\rightarrow$ & HCO & + & H$_2$O  &  2450.0 & & C & + & NH$_2$  & $\rightarrow$ & HNC & + & H  &  0.0 \\
H & + & C$_4$  & $\rightarrow$ & C$_4$H &  &   &  0.0 & & C & + & CH$_3$  & $\rightarrow$ & C$_2$H$_3$ &  &   &  0.0 \\
H & + & C$_4$H  & $\rightarrow$ & HC$_4$H &  &   &  0.0 & & C & + & O  & $\rightarrow$ & CO &  &   &  0.0 \\
H & + & HC$_4$H  & $\rightarrow$ & C$_4$H$_3$ &  &   &  1210.0 & & C & + & O$_2$  & $\rightarrow$ & CO & + & O  &  0.0 \\
H & + & C$_4$H$_3$  & $\rightarrow$ & CH$_2$CHCCH &  &   &  0.0 & & C & + & OH  & $\rightarrow$ & CO & + & H  &  0.0 \\
H & + & C$_3$N  & $\rightarrow$ & HC$_3$N &  &   &  0.0 & & C & + & SO  & $\rightarrow$ & CO & + & S  &  0.0 \\
H & + & HCOOCH$_3$  & $\rightarrow$ & CH$_3$OH & + & HCO  &  2450.0 & & C & + & C$_2$  & $\rightarrow$ & C$_3$ &  &   &  0.0 \\
H & + & OCS  & $\rightarrow$ & CO & + & HS  &  0.0 & & C & + & C$_2$H  & $\rightarrow$ & C$_3$H &  &   &  0.0 \\
H & + & SO$_2$  & $\rightarrow$ & O$_2$ & + & HS  &  0.0 & & C & + & CN  & $\rightarrow$ & C$_2$N &  &   &  0.0 \\
\cmidrule(lr){1-8} \cmidrule(lr){10-17} \\
    \end{tabular}%
    }
    \label{table:grainsurfacereactions}
\end{table*}

\begin{table*}
    \contcaption{Grain-surface reactions included in the chemical network. Each reaction is included following both the Langmuir-Hinshelwood and Eley-Rideal mechanism. The assumed activation barrier of the reaction $E_\mathrm{a}$ (K) is also listed.
    }
    \resizebox{0.9\textwidth}{!}{%
    \centering
    \begin{tabular}{l l l l l l l l l l l l l l l l l l l}
    \hline  
    \multicolumn{3}{c}{Reactants} & & \multicolumn{3}{c}{Products} & $E_\mathrm{a}$ [K] & & \multicolumn{3}{c}{Reactants} & & \multicolumn{3}{c}{Products} & $E_\mathrm{a}$ [K]\\
    \cmidrule(lr){1-8} \cmidrule(lr){10-17}
C & + & C$_2$H$_3$  & $\rightarrow$ & CH$_2$CCH &  &   &  0.0 & & CH & + & HNO  & $\rightarrow$ & NO & + & CH$_2$  &  0.0 \\
C & + & HS  & $\rightarrow$ & CS & + & H  &  0.0 & & CH & + & CH  & $\rightarrow$ & C$_2$H$_2$ &  &   &  0.0 \\
C & + & S  & $\rightarrow$ & CS &  &   &  0.0 & & CH & + & NH  & $\rightarrow$ & HNC & + & H  &  0.0 \\
C & + & C$_3$  & $\rightarrow$ & C$_4$ &  &   &  0.0 & & CH & + & NH  & $\rightarrow$ & HCN & + & H  &  0.0 \\
C & + & C$_3$H  & $\rightarrow$ & C$_4$H &  &   &  0.0 & & CH & + & NO  & $\rightarrow$ & HCN & + & O  &  0.0 \\
C & + & C$_2$N  & $\rightarrow$ & C$_3$N &  &   &  0.0 & & CH & + & CH$_2$  & $\rightarrow$ & C$_2$H$_3$ &  &   &  0.0 \\
C & + & C$_2$O  & $\rightarrow$ & C$_3$O &  &   &  0.0 & & CH & + & NH$_2$  & $\rightarrow$ & CH$_2$NH &  &   &  0.0 \\
C & + & C$_4$  & $\rightarrow$ & C$_5$ &  &   &  0.0 & & CH & + & CH$_3$  & $\rightarrow$ & C$_2$H$_4$ &  &   &  0.0 \\
C & + & C$_4$H  & $\rightarrow$ & C$_5$H &  &   &  0.0 & & CH & + & O$_2$  & $\rightarrow$ & HCO & + & O  &  0.0 \\
C & + & C$_2$S  & $\rightarrow$ & C$_3$S &  &   &  0.0 & & CH & + & C$_2$  & $\rightarrow$ & C$_3$H &  &   &  0.0 \\
C & + & C$_5$  & $\rightarrow$ & C$_6$ &  &   &  0.0 & & CH & + & C$_2$H  & $\rightarrow$ & C$_3$H$_2$ &  &   &  0.0 \\
C & + & C$_5$H  & $\rightarrow$ & C$_6$H &  &   &  0.0 & & CH & + & C$_2$H$_3$  & $\rightarrow$ & CH$_2$CCH$_2$ &  &   &  0.0 \\
C & + & C$_6$  & $\rightarrow$ & C$_7$ &  &   &  0.0 & & CH & + & C$_2$H$_3$  & $\rightarrow$ & CH$_3$CCH &  &   &  0.0 \\
C & + & C$_6$H  & $\rightarrow$ & C$_7$H &  &   &  0.0 & & CH & + & C$_3$  & $\rightarrow$ & C$_4$H &  &   &  0.0 \\
C & + & C$_7$  & $\rightarrow$ & C$_8$ &  &   &  0.0 & & CH & + & C$_3$H  & $\rightarrow$ & HC$_4$H &  &   &  0.0 \\
C & + & C$_7$H  & $\rightarrow$ & C$_8$H &  &   &  0.0 & & CH & + & C$_4$  & $\rightarrow$ & C$_5$H &  &   &  0.0 \\
C & + & C$_8$  & $\rightarrow$ & C$_9$ &  &   &  0.0 & & CH & + & C$_4$H  & $\rightarrow$ & C$_5$H$_2$ &  &   &  0.0 \\
C & + & C$_8$H  & $\rightarrow$ & C$_9$H &  &   &  0.0 & & CH & + & C$_5$  & $\rightarrow$ & C$_6$H &  &   &  0.0 \\
C & + & C$_9$  & $\rightarrow$ & C$_{10}$ &  &   &  0.0 & & CH & + & C$_5$H  & $\rightarrow$ & C$_6$H$_2$ &  &   &  0.0 \\
C & + & C$_9$H  & $\rightarrow$ & C$_{10}$H &  &   &  0.0 & & CH & + & C$_6$  & $\rightarrow$ & C$_7$H &  &   &  0.0 \\
C & + & C$_{10}$  & $\rightarrow$ & C11 &  &   &  0.0 & & CH & + & C$_6$H  & $\rightarrow$ & C$_7$H$_2$ &  &   &  0.0 \\
O & + & HCO  & $\rightarrow$ & CO & + & OH  &  0.0 & & CH & + & C$_7$  & $\rightarrow$ & C$_8$H &  &   &  0.0 \\
O & + & HNO  & $\rightarrow$ & NO & + & OH  &  0.0 & & CH & + & C$_7$H  & $\rightarrow$ & C$_8$H$_2$ &  &   &  0.0 \\
O & + & O$_2$H  & $\rightarrow$ & O$_2$ & + & OH  &  0.0 & & CH & + & C$_8$  & $\rightarrow$ & C$_9$H &  &   &  0.0 \\
O & + & CH  & $\rightarrow$ & HCO &  &   &  0.0 & & CH & + & C$_8$H  & $\rightarrow$ & C$_9$H$_2$ &  &   &  0.0 \\
O & + & CH$_2$  & $\rightarrow$ & H$_2$CO &  &   &  0.0 & & CH & + & C$_9$H  & $\rightarrow$ & C$_{10}$H$_2$ &  &   &  0.0 \\
O & + & NH  & $\rightarrow$ & HNO &  &   &  0.0 & & OH & + & CH$_3$CHO  & $\rightarrow$ & HCOOH & + & CH$_3$  &  2400.0 \\
O & + & NH$_2$  & $\rightarrow$ & HNO & + & H  &  0.0 & & OH & + & CH$_3$CHO  & $\rightarrow$ & CH$_3$OH & + & HCO  &  2400.0 \\
O & + & NS  & $\rightarrow$ & NO & + & S  &  0.0 & & OH & + & H$_2$CO  & $\rightarrow$ & HCO & + & H$_2$O  &  2850.0 \\
O & + & O  & $\rightarrow$ & O$_2$ &  &   &  0.0 & & OH & + & H$_2$CO  & $\rightarrow$ & HCOOH & + & H  &  2850.0 \\
O & + & OH  & $\rightarrow$ & O$_2$H &  &   &  0.0 & & OH & + & CH$_3$  & $\rightarrow$ & CH$_3$OH &  &   &  0.0 \\
O & + & C$_2$  & $\rightarrow$ & C$_2$O &  &   &  0.0 & & OH & + & OH  & $\rightarrow$ & H$_2$O$_2$ &  &   &  0.0 \\
O & + & CN  & $\rightarrow$ & OCN &  &   &  0.0 & & OH & + & OH  & $\rightarrow$ & H$_2$O & + & O  &  0.0 \\
O & + & CO  & $\rightarrow$ & CO$_2$ &  &   &  1000.0 & & OH & + & CO  & $\rightarrow$ & CO$_2$ & + & H  &  400.0 \\
O & + & HCO  & $\rightarrow$ & CO$_2$ & + & H  &  0.0 & & OH & + & HCO  & $\rightarrow$ & HCOOH &  &   &  0.0 \\
O & + & HS  & $\rightarrow$ & SO & + & H  &  0.0 & & NH & + & H$_2$CO  & $\rightarrow$ & HCO & + & NH$_2$  &  2850.0 \\
O & + & S  & $\rightarrow$ & SO &  &   &  0.0 & & NH & + & CH$_2$  & $\rightarrow$ & CH$_2$NH &  &   &  0.0 \\
O & + & C$_3$  & $\rightarrow$ & C$_3$O &  &   &  0.0 & & NH & + & NH  & $\rightarrow$ & N2 & + & H$_2$  &  0.0 \\
O & + & CS  & $\rightarrow$ & OCS &  &   &  0.0 & & NH & + & NO  & $\rightarrow$ & N2 & + & O  &  0.0 \\
O & + & SO  & $\rightarrow$ & SO$_2$ &  &   &  0.0 & & NH & + & CO  & $\rightarrow$ & HNCO &  &   &  1500.0 \\
N & + & O$_2$H  & $\rightarrow$ & O$_2$ & + & NH  &  0.0 & & CH$_2$ & + & O$_2$  & $\rightarrow$ & H$_2$CO & + & O  &  0.0 \\
N & + & CH  & $\rightarrow$ & HCN &  &   &  0.0 & & CH$_2$ & + & CN  & $\rightarrow$ & CH$_2$CN &  &   &  0.0 \\
N & + & CH$_2$  & $\rightarrow$ & H$_2$CN &  &   &  0.0 & & NH$_2$ & + & NO  & $\rightarrow$ & H$_2$O & + & N2  &  0.0 \\
N & + & N  & $\rightarrow$ & N2 &  &   &  0.0 & & CH$_2$ & + & CH$_2$  & $\rightarrow$ & C$_2$H$_4$ &  &   &  0.0 \\
N & + & NH  & $\rightarrow$ & N2 & + & H  &  0.0 & & CH$_2$ & + & CH$_3$  & $\rightarrow$ & C$_2$H5 &  &   &  0.0 \\
N & + & NS  & $\rightarrow$ & N2 & + & S  &  0.0 & & CH$_2$ & + & HNO  & $\rightarrow$ & NO & + & CH$_3$  &  0.0 \\
N & + & CH$_3$  & $\rightarrow$ & CH$_2$NH &  &   &  0.0 & & NH$_2$ & + & H$_2$CO  & $\rightarrow$ & NH$_3$ & + & HCO  &  2850.0 \\
N & + & O  & $\rightarrow$ & NO &  &   &  0.0 & & CH$_3$ & + & CN  & $\rightarrow$ & CH$_3$CN &  &   &  0.0 \\
N & + & C$_2$  & $\rightarrow$ & C$_2$N &  &   &  0.0 & & CH$_3$ & + & HNO  & $\rightarrow$ & CH$_4$ & + & NO  &  0.0 \\
N & + & HS  & $\rightarrow$ & NS & + & H  &  0.0 & & CH$_3$ & + & CH$_3$  & $\rightarrow$ & CH$_3$CH$_3$ &  &   &  0.0 \\
N & + & S  & $\rightarrow$ & NS &  &   &  0.0 & & CH$_3$ & + & HCO  & $\rightarrow$ & CH$_3$CHO &  &   &  0.0 \\
N & + & C$_3$  & $\rightarrow$ & C$_3$N &  &   &  0.0 & & CH$_3$ & + & H$_2$CO  & $\rightarrow$ & CH$_4$ & + & HCO  &  4450.0 \\
N & + & C$_3$H  & $\rightarrow$ & HC$_3$N &  &   &  0.0 & & CH$_3$ & + & CH$_3$CHO  & $\rightarrow$ & CH$_3$CH$_3$ & + & HCO  &  2400.0 \\
N & + & C$_5$  & $\rightarrow$ & C$_5$N &  &   &  0.0 & & CH$_3$ & + & C$_3$N  & $\rightarrow$ & CH$_3$C$_3$N &  &   &  0.0 \\
N & + & C$_5$H  & $\rightarrow$ & HC$_5$N &  &   &  0.0 & & CH$_3$ & + & C$_5$N  & $\rightarrow$ & CH$_3$C$_5$N &  &   &  0.0 \\
N & + & C$_7$  & $\rightarrow$ & C$_7$N &  &   &  0.0 & & CH$_3$ & + & C$_7$N  & $\rightarrow$ & CH$_3$C$_7$N &  &   &  0.0 \\
N & + & C$_7$H  & $\rightarrow$ & HC$_7$N &  &   &  0.0 & & C$_2$H & + & CH$_4$  & $\rightarrow$ & C$_2$H$_2$ & + & CH$_3$  &  250.0 \\
N & + & C$_9$  & $\rightarrow$ & C$_9$N &  &   &  0.0 & & & &  &  &  &  &  &  \\
N & + & C$_9$H  & $\rightarrow$ & HC$_9$N &  &   &  0.0 & & & &  &  &  &  &  &  \\
S & + & CH  & $\rightarrow$ & HCS &  &   &  0.0 & & & &  &  &  &  &  &  \\
S & + & CH$_3$  & $\rightarrow$ & H$_2$CS & + & H  &  0.0 & & & &  &  &  &  &  &  \\
S & + & NH  & $\rightarrow$ & NS & + & H  &  0.0 & & & &  &  &  &  &  &  \\
S & + & CO  & $\rightarrow$ & OCS &  &   &  0.0 & & & &  &  &  &  &  &  \\
\cmidrule(lr){1-8} \cmidrule(lr){10-17} \\
    \end{tabular}%
    }
    \footnotesize
    \begin{flushleft}
    {{\bf References} -- All activation barriers are those adopted in the grain-surface network compiled by \citet{Garrod2008}, except for the following:
    (i) \ce{H + HCN} and \ce{H2 + C}, a high barrier is adopted, constrained by modelling  to limit the formation of \ce{CH3NH2} and \ce{CH4} ice, respectively, at 10~K \citep{Penteado2017}; (ii) \ce{OH + CO}, from \citet{Garrod2011} and  \citet{Chen2005}; (iii) \ce{H + O2}, \ce{H + H2O2}, \ce{H2 + O}, \ce{H2 + O2H}, from \citet{Lamberts2013}.
    }
    \end{flushleft}
    \label{table:grainsurfacereactions-cont}
\end{table*}

\section{Fitted MCMAX dust temperature profiles}				\label{sect:app:tdustprofiles}

\begin{figure*}
 \includegraphics[width=0.75\textwidth]{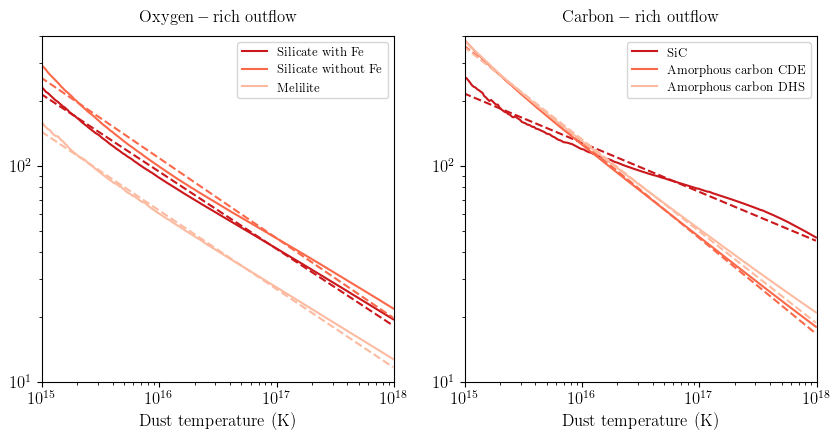}
 \caption{Dust temperature profiles calculated by MCMax (solid lines) together with the fitted power law (Eq.~\ref{eq:tdust}, dashed lines) for the highest density outflow with $\dot{M} = 10^{-5}$ M$_\odot$ yr$^{-1}$ and $v_\infty = 5$ km s$^{-1}$. Left panel: O-rich dust species. Right panel: C-rich dust species. The parameters of the fitted power law are listed in Table \ref{table:temperaturedust}.
 }
 \label{fig:tdustprofiles}
\end{figure*}

Fig. \ref{fig:tdustprofiles} shows the dust temperature profile calculated using MCMax together with the fit used in the model (Table \ref{table:temperaturedust}) for the highest density outflow with $\dot{M} = 10^{-5}$ M$_\odot$ yr$^{-1}$ and $v_\infty = 5$ km s$^{-1}$.

\section{Additional abundance profiles}				\label{sect:app:extrafigs}

\begin{figure*}
 \includegraphics[width=\textwidth]{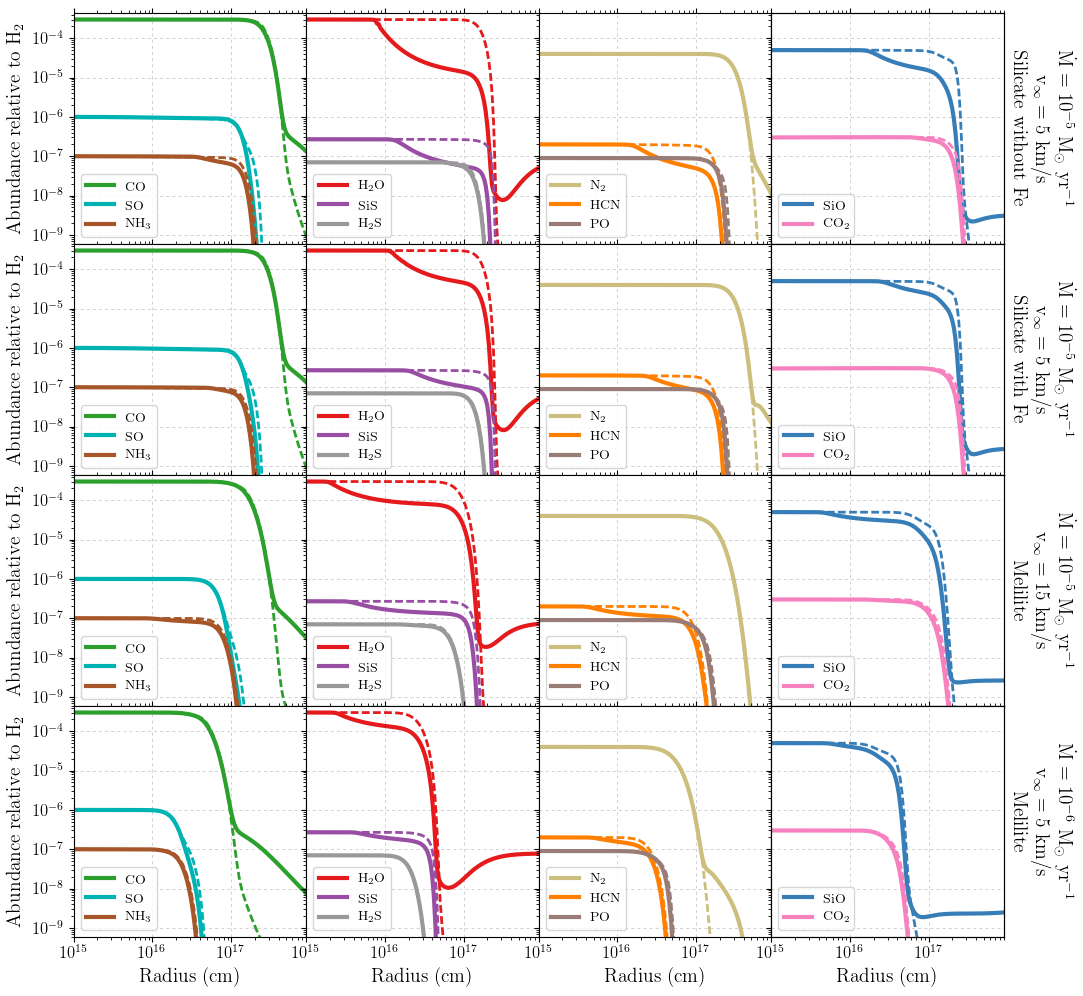}
 \caption{Abundance profiles of the O-rich parent species without dust-gas chemistry (dashed lines) and with dust-gas chemistry (solid lines) for the different dust types and densities as shown in Fig. \ref{fig:orich-depletion-h2o}. Complementary to Fig. \ref{fig:orich-maxdepletion-parents}. 
 }
 \label{fig:orich-parents-appendix}
\end{figure*}

\begin{figure*}
 \includegraphics[width=\textwidth]{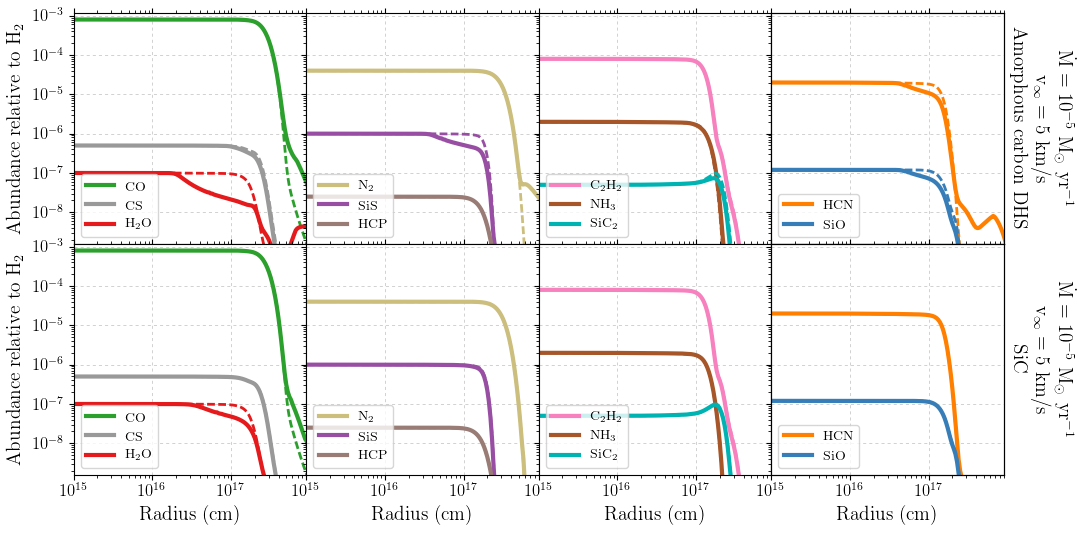}
 \caption{Abundance profiles of the C-rich parent species without dust-gas chemistry (dashed lines) and with dust-gas chemistry (solid lines) for the different dust types in an outflow with $\dot{M} = 10^{-5}$ M$_\odot$ yr$^{-1}$ and $v_\infty = 5$ km s$^{-1}$. Complementary to Fig. \ref{fig:orich-maxdepletion-parents}. 
 }
 \label{fig:crich-parents-appendix}
\end{figure*}

Fig. \ref{fig:orich-parents-appendix} shows the behaviour of the O-rich parent species in outflows of different overall density and for different dust grain types, complementary to Fig. \ref{fig:orich-maxdepletion-parents}.
Similarly, Fig. \ref{fig:crich-parents-appendix} shows the behaviour of the C-rich parent species in an outflow with $\dot{M} = 10^{-5}$ M$_\odot$ yr$^{-1}$ and $v_\infty = 5$ km s$^{-1}$ for different dust types, complementary to Fig. \ref{fig:crich-maxdepletion-parents}.

\end{appendix}



\bibliographystyle{mnras}
\bibliography{chemistry}






\bsp	
\label{lastpage}
\end{document}